\documentclass{article}
\usepackage{amssymb}

\usepackage{amsmath}
\usepackage{graphicx}

\newtheorem{theorem}{Theorem}

\newtheorem{lemma}[theorem]{Lemma}

\newtheorem{proposition}[theorem]{Proposition}

\input{tcilatex}

\begin{document}

\begin{center}
{\Large \textbf{\ Moduli space intersection duality between Regge surfaces}}

{\Large \textbf{\ and }}

{\Large \textbf{2D dynamical triangulations}}

\bigskip

{\large \textsl{M. Carfora}$\,^{b,}$\footnote{{\large email
mauro.carfora@pv.infn.it}},} \vspace{24pt}

{\large \textsl{A. Marzuoli}$\,^{b,}$\footnote{{\large email
annalisa.marzuoli@pv.infn.it}},} \vspace{24pt}

{\large \textsl{P. Villani}$\,^{b,}$\footnote{{\large email
paolo.villani@pv.infn.it}},} \vspace{24pt}

$^{b}$~Dipartimento di Fisica Nucleare e Teorica,

Universit\`{a} degli Studi di Pavia, \\[0pt]
via A. Bassi 6, I-27100 Pavia, Italy, \\[0pt]
and\\[0pt]
Istituto Nazionale di Fisica Nucleare, Sezione di Pavia, \\[0pt]
via A. Bassi 6, I-27100 Pavia, Italy

\textbf{Abstract}
\end{center}

Deformation theory for $2$-dimensional dynamical triangulations with $N_{0}$
vertices is discussed by exploiting the geometry of the moduli space of
Euclidean polygons. Such an analysis provides an explicit connection among
Regge surfaces, dynamical triangulations theory and the Witten-Kontsevich
model. In particular we show that a natural set of Regge measures and a
triangulation counting of relevance for dynamical triangulations are
directly connected with intersection theory over the (compactified) moduli
space \ $\overline{\frak{M}}_{g},_{N_{0}}$ of genus $g$ Riemann surfaces
with $N_{0}$ punctures. The Regge measures in question provide volumes of
the open strata of \ $\overline{\frak{M}}_{g},_{N_{0}}$. From the physical
point of view, the arguments presented here offer evidence that quantum
Regge calculus and dynamical triangulations are related by a form of
topological $S$-duality.

\medskip



\newpage

\section{\protect\bigskip Introduction}

The successful analysis of two-dimensional quantum gravity can be partly
traced back to the quantum geometry of piecewise-linear surfaces.\ We
discuss in this paper \ a full-fledged geometrical interplay among two of
the main characters of such a theory:\ dynamical triangulations, and Regge
surfaces. We trace a logical path from Riemann moduli theory to
triangulations of surfaces that puts clearly to the fore the deep
interconnections between dynamical triangulations and Regge calculus, a
connection that goes far beyhond the few algorithmic elements of
piecewise-linear geometry that are common to both theories. We show
explicitly that Dynamical triangulations and Regge surfaces with $N_{0}$
vertices are, in a mathematically well-defined sense, dual under a natural
pairing in the (compactified) moduli space of genus $g$ Riemann surfaces
with $N_{0}$ punctures $\overline{\frak{M}}_{g},_{N_{0}}$, a duality related
to the Witten-Kontsevich intersection numbers $\left\langle \tau _{\delta
_{1}}...\tau _{\delta _{N_{0}}}\right\rangle $, [1]. Dynamical
triangulations have a natural interpretation in terms of open strata of $\ 
\overline{\frak{M}}_{g},_{N_{0}}$, \ i.e., to a dynamical triangulation we
can associate a generic region of the moduli space. This is not surprising
since is just a restatement in the language of dynamical triangulations of
the well-known combinatorial parametrization of $\overline{\frak{M}}%
_{g},_{N_{0}}$ in terms of ribbon graph theory (i.e., of graphs which can be
drawn on surfaces), exploited by Kontsevich. In our a framework what is new
is the role of Regge triangulations. They naturally appear upon deforming
dynamically triangulated surfaces, and can be related to the geometry of the
moduli space of (similarity classes of) polygons in the Euclidean plane.
Such a moduli space carries a $2$-form that induces a natural volume on the
set of Regge surfaces. The structure of the resulting Regge measure is
surprisingly elementary, yet it is deeply connected with the geometry of $\ 
\overline{\frak{M}}_{g},_{N_{0}}$ and provides a natural mean of computing
the volume of the open strata in $\overline{\frak{M}}_{g},_{N_{0}}$ labelled
by the given dynamical triangulations. The actual computation of such a
Regge measure is directly related to the volume of a simplex in a Euclidean
space whose dimensions are given in terms of \ the number of vertices
(punctures) $N_{0}$ and of the genus $g$ of the triangulated surface. We
leave to a subsequent paper the difficult technical aspects involved in the
explicit computation of such a volume. Perhaps what it is more important to
stress here is the fact that our analysis provides evidence that
two-dimensional quantum Regge calculus \ and dynamical triangulations theory
are related by a form of topological $S$-duality. Such a duality comes about
by observing that the set of Regge-like measures mentioned above and the
entropy of dynamical triangulations play against each other, \emph{a' la
Peierls}, so as to generate the Witten-Kontsevich intersection numbers.

\bigskip

Coming to the structure of the paper, we tried, as far as possible, to
provide a complete and self-contained explanation to the connection between
dynamical triangulations and Regge surfaces. Many facts that we present here
in detail are often well-known in disguised form to the specialist in moduli
theory, but the relation with triangulations is so subtle that we have
preferred to be quite explicit. The path we follow is partly related to the
geometrical setup of the Witten-Kontsevich model and the combinatorial
parametrization of $\overline{\frak{M}}_{g},_{N_{0}}$\ in terms of ribbon
graphs theory. However, the different emphasis we put here on Regge calculus
gives to our formulation a different flavor which leads to a simple
understanding of the interplay between Regge surfaces, dynamical
triangulations, and moduli theory.

\section{\protect\bigskip Singular Euclidean structures}

Let $T$ denote a $n$-dimensional simplicial complex with underlying
polyhedron $|T|$ and $f$- vector $(N_{0}(T),N_{1}(T),\ldots ,N_{n}(T))$,
where $N_{i}(T)\in \mathbb{N}$ is the number of $i$-dimensional sub-
simplices $\sigma ^{i}$ of $T$. If we consider the (first) barycentric
subdivision of \ $T$, then the closed stars, in such a subdivision, of the
vertices of the original triangulation $T$ form a collection of $n$-cells
characterzing\ the polytope $P$ barycentrically dual to $T$. A Regge
triangulation of a $n$-dimensional PL manifold $M$, with boundary $\partial
M $, is a homeomorphism $|T_{l}|\rightarrow {M}\ $ where each face of $T$ \
\ is geometrically realized by a rectilinear simplex of variable
edge-lengths $l(\sigma ^{1}(k))$\ of the appropriate dimension. A dynamical
triangulation $|T_{l=a}|\rightarrow {M}$ \ is \ a particolar case of a Regge
PL-manifold realized by rectilinear and equilateral simplices of edge-length 
$l(\sigma ^{1}(k))=$ $a$. Similarly, a rectilinear presentation $%
|P_{T_{L}}|\rightarrow {M}$ of the dual cell complex $P$ (with edge-lengths $%
L=L(l)$) characterizes the Regge polytope (and its rigid equilateral
specialization $|P_{T_{a}}|\rightarrow {M}$) \ baricentrically dual to $%
|T_{l}|\rightarrow {M}$. The metric structure of a Regge triangulation\ is
locally Euclidean everywhere except at the $(n-2)$ sub-simplices $\sigma
^{n-2}$, (the \textit{bones}), where the sum of the dihedral angles, $\theta
(\sigma ^{n})$, of the incident $\sigma ^{n}$'s is in excess (negative
curvature) or in defect (positive curvature) with respect to the $2\pi $
flatness constraint. The corresponding deficit angle $r$ is defined by $%
r=2\pi -\sum_{\sigma ^{n}}\theta (\sigma ^{n})$, where the summation is
extended to all $n$-dimensional simplices incident on the given bone $\sigma
^{n-2}$. If $K_{T}^{n-2}$ denotes the $(n-2)$-skeleton of $%
|T_{l}|\rightarrow {M}$, then \ $M\backslash {K_{T}^{n-2}}$ is a flat
Riemannian manifold, and any point in the interior of an $r$- simplex $%
\sigma ^{r}$ has a neighborhood homeomorphic to $B^{r}\times {C}(lk(\sigma
^{r}))$, where $B^{r}$ denotes the ball in $\mathbb{R}^{n}$ and ${C}%
(lk(\sigma ^{r}))$ is the cone over the link $lk(\sigma ^{r})$, (the product 
$lk(\sigma ^{r})\times \lbrack 0,1]$ with $lk(\sigma ^{r})\times \{1\}$
identified to a point), (recall that if we denote by $st(\sigma )$, (the
star of $\sigma $), the union of all simplices of which $\sigma $ is a face,
then $lk(\sigma ^{r})$ is the union of all faces $\sigma ^{f}$\ of the
simplices in $st(\sigma )$ such that $\sigma ^{f}\cap \sigma =\emptyset $).
For dynamical triangulations, the deficit angles are generated by the string
of integers, the \textit{curvature assignments}, $\{q(k)\}_{k=1}^{N_{n-2}}%
\in \mathbb{N}^{N_{n-2}(T)}$,\ \textit{viz.},

\begin{gather}
r(i)=2\pi -q(i)\arccos (1/n),\;\sigma ^{n-2}(i)\in M\backslash \partial M
\label{curvature} \\
=\pi -q(i)\arccos (1/n),\;\sigma ^{n-2}(i)\in \partial M  \notag
\end{gather}
where

\begin{gather}
q(\cdot ):\{\sigma ^{n-2}(k)\}_{k=1}^{N_{n-2}(T)}\rightarrow \mathbb{N}^{+} 
\notag \\
\sigma ^{n-2}(i)\longmapsto q(i)\doteq \#\{\sigma ^{n}(h)\bot \sigma
^{n-2}(i)\}
\end{gather}
provides the numbers of top-dimensional simplices incident on the $N_{n-2}$
distinct bones. \ Since each top-dimensional simplex has $\frac{1}{2}n(n+1)$
bones $\sigma ^{n-2}$, the set of integers $\{q(k)\geq 3\}_{k=1}^{N_{n-2}}$
\ is constrained by

\begin{equation}
\sum_{k}^{N_{n-2}}q(k)=\frac{1}{2}n(n+1)N_{n}=b(n,n-2)N_{n-2},
\label{vincolo}
\end{equation}
where $b(n,n-2)$\ is the average value of the curvature assignments \ $%
\{q(k)\}_{k=1}^{N_{n-2}}$. \ 

As recalled, a Regge triangulation $|T_{l}|\rightarrow M$ defines on the PL
manifold $M$ a polyhedral metric with conical singularities associated with
the bones $\{\sigma ^{n-2}(i)\}_{i=1}^{N_{n-2}(T)}$ of the triangulation,
but which is otherwise flat and smooth everywhere else. Such a metric has
important special features, in particular it induces on the PL manifold \ $M$
a geometrical structure which turns out to be a particular case of the
theory of \emph{Singular \ Euclidean Structure} (in the sense of M. Troyanov
[2], and W. Thurston [3]).\ Since each bone $\sigma ^{n-2}(k)$ is $(n-2)-$%
dimensional, a point in the interior of $\sigma ^{n-2}(k)$ has a
neighborhood in $|T_{L}|\rightarrow M$ \ which is homeomorphic to $%
B^{n-2}(k)\times B^{2}(k)$, where $B^{m}(k)$ is an $m-$dimensional
topological ball. Explicitly, the disc $B^{2}(k)\simeq $ $C|lk(\sigma
^{n-2}(k))|$ is the cone over the link of the bone. On such a disc we can
introduce a locally uniformizing complex coordinate $\zeta _{k}\in \mathbb{C}
$ in terms of which we can explicitly write down the singular (conformal)
Euclidean metric locally characterizing the singular Euclidean structure of $%
\ |T_{l}|\rightarrow M$, \emph{viz.}, 
\begin{equation}
ds_{(k)}^{2}\doteq e^{2u}\left| \zeta _{k}-\zeta _{k}(\sigma
^{n-2}(k))\right| ^{-2\left( \frac{r(k)}{2\pi }\right) }\left| d\zeta
_{k}\right| ^{2},  \label{cmetr}
\end{equation}
where \ $r(k)$ is given by (\ref{curvature}), and $u:B^{2}\rightarrow 
\mathbb{R}$ is a continuous function ($C^{2}$ on $B^{2}-\{\sigma ^{n-2}(k)\}$%
).\ \ Up to the presence of the conformal factor $e^{2u}$, we immediately
recognize in such an expression the metric of a Euclidean cone of total
angle $\theta (k)=r(k)-2\pi $. Notationally, the metric (\ref{cmetr}) and
its singular structure can be naturally summarized in a formal linear
combination of the bones $\{\sigma ^{n-2}(k)\}$ with coefficients given by
the corresponding deficit angles (normalized to $2\pi $), \emph{viz.}, in\
the \emph{real divisor }[2] 
\begin{gather}
Div(T)\doteq \sum_{k=1}^{N_{n-2}(T)}\left( -\frac{r(k)}{2\pi }\right) \sigma
^{n-2}(k)=\sum_{k=1}^{N_{n-2}(T\backslash \partial T)}\left( \frac{\theta (k)%
}{2\pi }-1\right) \sigma ^{n-2}(k) \\
+\sum_{h=1}^{N_{n-2}(\partial T)}\left( \frac{\theta (h)}{2\pi }-\frac{1}{2}%
\right) \sigma ^{n-2}(h)  \notag
\end{gather}
supported on the set of bones $\{\sigma ^{n-2}(i)\}_{i=1}^{N_{n-2}(T)}$.
Note that the degree of such a divisor, \ defined by 
\begin{gather}
\left| Div(T)\right| \doteq \sum_{k=1}^{N_{n-2}(T\backslash \partial
T)}\left( \frac{\theta (k)}{2\pi }-1\right) +\sum_{h=1}^{N_{n-2}(\partial
T)}\left( \frac{\theta (h)}{2\pi }-\frac{1}{2}\right) \\
\underset{Dyn\,Triang}{=}\left[ \frac{\arccos (1/n)}{2\pi }b(n,n-2)-1\right]
N_{n-2}(T)+\frac{1}{2}N_{n-2}(\partial T)  \notag
\end{gather}
is, for dynamical triangulations, a rewriting of the combinatorial
constraint (\ref{vincolo}); in such a sense, the pair $(|T_{l=a}|\rightarrow
M,Div(T))$, or shortly, $(T_{a},Div(T))$, encodes the datum of the
triangulation $|T_{l=a}|\rightarrow M$ and of a corresponding set of
curvature assignments $\{q(k)\}$ on the labelled bones $\{\sigma
^{n-2}(i)\}_{i=1}^{N_{n-2}(T)}$. \ \ 

We conclude this introductory section by recalling a few topological remarks
explicitly pertaining to $2$-dimensional dynamically triangulated surfaces $%
|T_{l=a}|\rightarrow M$. Since, for $n=2$, \ the average incidence of $%
|T_{l=a}|\rightarrow M$ is provided by 
\begin{equation}
b(n,n-2)|_{n=2}=6\left[ 1-\frac{\chi (M)}{N_{0}(T)}\right] ,  \label{avedue}
\end{equation}
where $\chi (M)$ denotes the Euler-Poincar\'{e} characteristic of the
surface, we get $\left| Div(T)\right| _{(n=2)}=-\chi (M)$. Thus, the real
divisor $\left| Div(T)\right| _{(n=2)}$ characterizes the Euler class of \
the pair\ $(T_{a},Div(T))$ and yields for a corresponding Gauss-Bonnet
formula. Explicitly, the Euler number associated with $(T_{a},Div(T))$ is
defined, [2], by

\begin{equation}
e(T_{a},Div(T))\doteq \chi (M)+|Div(T)\mathbf{|.}
\end{equation}
and the Gauss-Bonnet formula reads:

\begin{lemma}
(\textbf{Gauss-Bonnet for DT surfaces}) Let $(T_{a},Div(T))$ be a
dynamically triangulated surface with divisor 
\begin{equation}
Div(T)\doteq \sum_{k=1}^{N_{0}(T)}\left( \frac{\theta (k)}{2\pi }-1\right)
\sigma ^{0}(k),
\end{equation}
associated with the vertex incidences $\{\theta (k)\doteq q(k)\frac{\pi }{3}%
\}_{k=1}^{N_{0}(T)}$. Let $ds^{2}$ be the conformal metric (\ref{cmetr})
representing the divisor $Div(T)$ . Then 
\begin{equation}
\frac{1}{2\pi }\iint_{M}KdA=e(T_{a},Div(T)),
\end{equation}
where $K$\ and $dA$\ respectively are the curvature and the area element
corresponding to the metric $ds^{2}.$
\end{lemma}

Note that such a theorem holds for any singular Riemann surface $\Sigma $
described by a divisor $Div(\Sigma )$ and not just for dynamically
triangulated surfaces [2]. Since for a dynamical triangulation, we have $%
e(T_{a},Div(T))=0$, the Gauss-Bonnet formula implies

\begin{equation}
\frac{1}{2\pi }\iint_{M}KdA=0.
\end{equation}
Thus, a dynamical triangulation $|T_{l=a}|\rightarrow M$ naturally carries a
conformally flat structure. Clearly this is a rather obvious result, (since
the metric in $M-\{\sigma ^{0}(i)\}_{i=1}^{N_{0}(T)}$ is flat). However, it
admits a not-trivial converse (recently proved by M. Troyanov, but, in a
sense, going back to E. Picard) [2], [4]:

\begin{theorem}
(\textbf{Troyanov-Picard}) Let $(M,Div)$ be a singular Riemann surface with
a divisor such that $e(M,Div)=0$. Then there exists on $M$\ a unique (up to
homothety) conformally flat metric representing the divisor $Div$.
\end{theorem}

This result geometrically characterizes dynamical triangulations (and Regge
surfaces) as a particular case of the theory of singular Riemann surfaces,
and provides the rationale for understanding at a deeper level the
connection between triangulations and some aspects of surface theory of
relevance to $2D$ gravity. A well known example of such a connection is
afforded by the relation between the dual polygonalization of $%
|T|\rightarrow M$ and the space of complete punctured Riemann surfaces
exploited in the Witten-Kontsevich theory [1]. A further and strictly
related example is provided by the moduli space intersection pairing between
Regge surfaces and dynamical triangulations which is the main theme of our
paper.

\section{\protect\bigskip Regge surfaces and ribbon graphs}

In dimension $n=2$, the geometrical realization of the $1$-skeleton of \ $%
|P_{T_{L}}|\rightarrow {M}$ is a $3$-valent graph $\Gamma =(\{\varrho
^{o}(k)\},\{e^{1}(j)\})$ where the vertex set $\{\varrho
^{o}(k)\}_{k=1}^{N_{2}(T)}$ is identified with the barycenters of the
triangles $\{\sigma ^{o}(k)\}_{k=1}^{N_{2}(T)}\in |T_{l}|\rightarrow M$,
whereas each edge $e^{1}(j)\in \{e^{1}(j)\}_{j=1}^{N_{1}(T)}$ is generated
by two half-edges $e^{1}(j)^{+}$ and $e^{1}(j)^{-}$ joined through the
barycenters $\{W(h)\}_{h=1}^{N_{1}(T)}$ of the edges $\{\sigma ^{1}(h)\}$
belonging to the original triangulation $|T_{l}|\rightarrow M$. Thus, if we
formally introduce a degree-$2$ vertex at each middle point $%
\{W(h)\}_{h=1}^{N_{1}(T)}$, the actual graph naturally associated to the $1$%
-skeleton of \ $|P_{T_{L}}|\rightarrow {M}$ is 
\begin{equation}
\Gamma _{ref}=\left( \{\varrho ^{o}(k)\}\bigsqcup
\{W(h)\},\{e^{1}(j)^{+}\}\bigsqcup \{e^{1}(j)^{-}\}\right) ,
\end{equation}
the so called edge-refinement [5] of $\Gamma =(\{\varrho
^{o}(k)\},\{e^{1}(j)\})$. The relevance of such a notion stems from the
observation that the natural automorphism group $Aut(P_{L})$ of \ $%
|P_{T_{L}}|\rightarrow {M}$, (\emph{i.e.}, the set of bijective maps $\Gamma
=(\{\varrho ^{o}(k)\},\{e^{1}(j)\})\rightarrow \widetilde{\Gamma }=(%
\widetilde{\{\varrho ^{o}(k)\}},\widetilde{\{e^{1}(j)\}}$ preserving the
incidence relations defining the graph structure), is not the automorphism
group of $\Gamma $ but rather the (larger) automorphism group of its edge
refinement [5], \emph{i.e.}, 
\begin{equation}
Aut(P_{L})\doteq Aut(\Gamma _{ref}).
\end{equation}
The locally uniformizing complex coordinate $\zeta _{k}\in \mathbb{C}$ \ in
terms of which we can explicitly write down the singular Euclidean metric (%
\ref{cmetr}) around each vertex $\sigma ^{0}(k)\in $ $|T_{l}|\rightarrow M$,
provides a (counterclockwise) orientation in the $2$-cells of $%
|P_{T_{L}}|\rightarrow {M}$. Such an orientation gives rise to a cyclic
ordering on the set of half-edges $\{e^{1}(j)^{\pm }\}_{j=1}^{N_{1}(T)}$
incident on the vertices $\{\varrho ^{o}(k)\}_{k=1}^{N_{2}(T)}$. In
particular we assume that the half-edges $e^{1}(j)^{+}$ and $e^{1}(j+1)^{-}$
are incident to the vertex $\varrho ^{o}(j)$ in such a way that $%
e^{1}(j)^{+} $ precedes $e^{1}(j+1)^{-}$ with respect to the given cyclic
ordering. According to such remarks, the $1$-skeleton of \ $%
|P_{T_{L}}|\rightarrow {M}$ is a ribbon (or fat) graph [5], \emph{viz.}, a
graph $\Gamma $ together with a cyclic ordering on the set of half-edges
incident to each vertex of $\ \Gamma $. In general, the set of half-edges of
(the edge-refinement of) a graph\ does not carry any particular structure.
It is precisely the introduction of a cyclic ordering on such a set that
allows one to characterize\ a ribbon graph as a graph which can be drawn on (%
\emph{i.e.}, embedded into) an oriented surface. Conversely, any ribbon
graph $\Gamma $ characterizes an oriented surface $M(\Gamma )$ with boundary
possessing $\Gamma $ as a spine, (\emph{i.e.}, the inclusion $\Gamma
\hookrightarrow M(\Gamma )$ is a homotopy equivalence). In order to
construct $M(\Gamma )$ , recall that the half-edges incident to a generic
vertex $\varrho ^{o}(j)$ can be described by thin strips joined at the
vertex and with the cyclic ordering at $\varrho ^{o}(j)$ determining a
direction on the boundaries of the strip. According to such a description,
the surface $M(\Gamma )$ is generated by connecting the half-edges
coherently with their orientation, in such a way that $\partial M(\Gamma )$
is identified with the boundary of the ribbon graph $\Gamma $. Since this
latter boundary is oriented, we can attach an oriented disk $D^{2}(j)$ to
each boundary component of $\Gamma $. The resulting space $M$ is a compact
oriented topological surface whose genus $g(M)$ is given by 
\begin{equation}
2-2g(M)=|v(\Gamma )|-|e(\Gamma )|+|\partial (\Gamma )|,
\end{equation}
where $|v(\Gamma )|$, $|e(\Gamma )|$, and $|\partial (\Gamma )|$
respectively denote the number of vertices, edges and boundary components of 
$\Gamma $. Since we want to keep track of the fact that $\Gamma $ is
associated with a polytope $|P_{T_{L}}|\rightarrow {M}$ dual to a Regge
triangulation, it is natural to attach to the oriented boundaries $\partial
M(\Gamma )$ \ the length structure naturally associated with $%
|P_{T_{L}}|\rightarrow {M}$. Note that rather than the disks $D^{2}(j)$, we
may glue to the boundary components of the ribbon graph $\Gamma $ a
corresponding set of punctured disks $D_{punct}^{2}(j)$. The punctures can
be profitably identified with the vertices $\{\sigma ^{0}(k)\}$\ of the
Regge triangulation $|T_{l}|\rightarrow M$ which, upon barycentrical
dualization, gives rise to $|P_{T_{L}}|\rightarrow {M}$. In this way (the
edge-refinement of ) the $1$-skeleton of \ $|P_{T_{L}}|\rightarrow {M}$
acquires from the underlying Regge triangulation the structure of a metric
ribbon graph. In general, the set of all possible metrics on a (trivalent)
ribbon graph $\Gamma $ with given edge-set $e(\Gamma )$\ can be
characterized (see [5], definition 3.1) as a topological space homeomorphic
to $\mathbb{R}_{+}^{|e(\Gamma )|}$, ($|e(\Gamma )|$ denoting the number of
edges in $e(\Gamma )$), topologized by the \ standard $\epsilon $%
-neighborhoods \ $U_{\epsilon }\subset $ $\mathbb{R}_{+}^{|e(\Gamma )|}$. On
such a space there is a natural action of $Aut(\Gamma )$, the automorphism
group of $\Gamma $ defined by the homomorphism $Aut(\Gamma )\rightarrow 
\frak{G}_{e(\Gamma )}$ where $\frak{G}_{e(\Gamma )}$ denotes the symmetric
group over $|e(\Gamma )|$ elements. Thus, the resulting space $\mathbb{R}%
_{+}^{|e(\Gamma )|}/Aut(\Gamma )$ is a differentiable orbifold, (the
quotient of a manifold by a finite group). Let 
\begin{equation}
Aut_{\partial }(P_{L})\subset Aut(P_{L}),
\end{equation}
denote the subgroup of ribbon graph automorphisms of \ the (trivalent) $1$%
-skeleton $\Gamma $ of \ $|P_{T_{L}}|\rightarrow {M}$\ that preserve the
(labeling of the) boundary components of $\Gamma $. Then, the space of \ $1$%
-skeletons of \ Regge polytopes\ $|P_{T_{L}}|\rightarrow {M}$, with $%
N_{0}(T) $ labelled boundary components, on a surface $M$ of genus $g$ can
be defined by [5] 
\begin{equation}
RGP_{g,N}^{met}=\bigsqcup_{\Gamma \in RGB_{g,N}}\frac{\mathbb{R}%
_{+}^{|e(\Gamma )|}}{Aut_{\partial }(P_{L})},  \label{DTorb}
\end{equation}
where the disjoint union is over the subset of all trivalent ribbon graphs
(with labelled boundaries) satisfying the topological condition $%
2-2g-N_{0}(T)<0$, and which are dual to triangulations. In this connection,
it is worth noticing that not all trivalent ribbon graphs are dual to
regular triangulations (\emph{e.g.}, degenerate triangulations with \emph{%
pockets}, where two triangles are incident on a vertex, give rise to
trivalent dual ribbon graphs with loops, \emph{i.e.} Regge polytopes
containing $2$-gons). Our analysis can be extended to such degenerate case
as well, however the Regge measure (\ref{tildeform}) we will use eliminates
such singular configurations. Also the set $RGP_{g,N}^{met}$ has a natural
structure of differentiable orbifold, locally modelled on a stratified space
constructed from the components $\mathbb{R}_{+}^{|e(\Gamma )|}/Aut_{\partial
}(P_{L})$ by means of a (Whitehead) expansion and collapse procedure for
ribbon graphs, (see [5] theorems 3.3, 3.4, and 3.5), which basically amounts
to collapsing edges and coalescing vertices. This subject has been developed
to a high degree of sophistication in [5], and we do not pursue it here any
further. As a crude argument, suffices it to say that since in Regge
calculus we work at fixed connectivity (\emph{i.e.}, the adjacency matrix of
the triangulation $|P_{T_{L}}|\rightarrow {M}$ is fixed a priori), the
relevant \emph{model space} for the set of $1$-skeletons of \ Regge
polytopes\ $|P_{T_{L}}|\rightarrow {M}$\ is the differential orbifold 
\begin{equation}
\frac{\mathbb{R}_{+}^{|e(\Gamma )|}}{Aut_{\partial }(P_{L})},  \label{ReOrb}
\end{equation}
namely a rational cell of \ the most general orbifold (\ref{DTorb}). Note
that the role of the automorphism group $Aut_{\partial }(P_{L})$ is highly
non-trivial, and the orbifold structure of (\ref{ReOrb}) has a deep impact
on the topology of the configuration space of all Regge triangulations, and
one cannot simply take $\mathbb{R}_{+}^{|e(\Gamma )|}$ as a local model for
the metric geometry of a Regge polytope\ $|P_{T_{L}}|\rightarrow {M}$. For
dynamical triangulations the adjacency matrix varies, and we must use the
general orbifold (\ref{DTorb}) as a model of all the rational cells
comprising the possible metric geometries of the rigid polytopes\ $%
|P_{T_{a}}|\rightarrow {M}$.

\subsection{Local deformations of Regge surfaces}

The edge-refinement of the $1$-skeleton of \ $|P_{T_{L}}|\rightarrow {M}$
comes to the fore also in discussing the explicit connection \ between the
metric geometry of a Regge surface $|T_{l}|\rightarrow M$ and the metric
structure of \ the corresponding barycentrically dual complex $%
|P_{T_{L}}|\rightarrow {M}$. As a rule, this latter aspect is not emphasized
in discussing Regge surfaces, (and it is rather trivial for dynamical
triangulations owing to the equilateral constraint), but for later use we
need to discuss it here in some detail. To this end, let us fix our
attention on the generic vertex $\sigma ^{0}(k)\in $ $|T_{l}|\rightarrow M$,
\ and let us path-order the $q(k)$ vertices of the link $lk(\sigma ^{0}(k))$%
. If we denote such a collection of ordered vertices by $V_{\alpha }(k)$, $%
\alpha =1,...,q(k)$, then by splitting open the star $st(\sigma ^{0}(k))$
along the edge connecting $\sigma ^{0}(k)$ to $V_{1}(k)$ we generate a $2$%
-dimensional Euclidean simplicial complex with $q(k)$ triangles 
\begin{equation}
\{\Delta _{\alpha }(k)\doteq (V_{\alpha }(k),\sigma ^{0}(k),V_{\alpha
+1}(k))\}_{\alpha =1}^{q(k)}
\end{equation}
all incident on the common vertex $\sigma ^{0}(k)$ and such that $\Delta
_{\alpha }(k)$ shares with the adjacent triangle $\Delta _{\alpha +1}(k)$
the edge $V_{\alpha +1}(k)$, (with $\alpha +1=1\,\func{mod}q(k)$). In each
triangle $\Delta _{\alpha }(k)$ we can introduce a coordinate system
centered at $O\equiv $ $\sigma ^{0}(k)$ with unit basis vectors $%
\overrightarrow{e}_{1}(\alpha )$ and $\overrightarrow{e}_{2}(\alpha )$,
respectively directed along the edges $OV_{\alpha }(k)$ and $OV_{\alpha
+1}(k)$. Thus, 
\begin{gather}
\overrightarrow{OV_{\alpha }(k)}=l_{\alpha }(k)\overrightarrow{v}_{1}(\alpha
), \\
\overrightarrow{OV_{\alpha +1}(k)}=l_{\alpha +1}(k)\overrightarrow{v}%
_{2}(\alpha ),  \notag
\end{gather}
where $l_{\alpha }(k)$ and $l_{\alpha +1}(k)$ denote the respective
edge-lengths in the Regge triangulation $|T_{l}|\rightarrow M$ . Note that 
\begin{gather}
\overrightarrow{V_{\alpha }(k)V_{\alpha +1}(k)}=\overrightarrow{OV_{\alpha
+1}(k)}-\overrightarrow{OV_{\alpha }(k)}, \\
\overrightarrow{v}_{1}(\alpha )\cdot \overrightarrow{v}_{2}(\alpha )=\cos
\;\theta _{\alpha ,\alpha +1}(k),  \notag
\end{gather}
where $\theta _{\alpha ,\alpha +1}(k)$ is the (dihedral) angle at $\sigma
^{0}(k)$ between the edges $\overrightarrow{OV_{\alpha }(k)}$ and $%
\overrightarrow{OV_{\alpha +1}(k)}$. The quantity 
\begin{equation}
2\pi -\sum_{\alpha =1}^{q(k)}\theta _{\alpha ,\alpha +1}(k)
\end{equation}
is the deficit angle of the Regge triangulation $|T_{l}|\rightarrow M$ \ at
the given vertex $\sigma ^{0}(k)$. In such a setting, the barycenter \ $%
\varrho _{\alpha }^{o}(k)$ of the triangle $\Delta _{\alpha }(k)$ is given
by 
\begin{equation}
\overrightarrow{O\varrho _{\alpha }^{o}(k)}\doteq \frac{1}{3}\left[ 
\overrightarrow{OV_{\alpha }(k)}+\overrightarrow{OV_{\alpha +1}(k)}\right] ,
\end{equation}
whereas the vector $\overrightarrow{\varrho _{\alpha }^{o}(k)W_{\alpha +1}(k)%
}$ connecting the barycenter with the midpoint $W_{\alpha +1}(k)$ of the
edge $OV_{\alpha +1}(k)$ is provided by 
\begin{gather}
\overrightarrow{\varrho _{\alpha }^{o}(k)W_{\alpha +1}(k)}=\frac{1}{2}%
\overrightarrow{OV_{\alpha +1}(k)}-\overrightarrow{O\varrho _{\alpha }^{o}(k)%
} \\
=\frac{1}{6}\overrightarrow{OV_{\alpha +1}(k)}-\frac{1}{3}\overrightarrow{%
OV_{\alpha }(k)}.  \notag
\end{gather}
In particular, we get 
\begin{gather}
L_{\alpha }^{+}(k)\equiv \left| \overrightarrow{\varrho _{\alpha
}^{o}(k)W_{\alpha +1}(k)}\right|  \label{plushalf} \\
=\left[ \frac{1}{36}l_{\alpha +1}^{2}(k)+\frac{1}{9}l_{\alpha }^{2}(k)-\frac{%
1}{9}l_{\alpha +1}(k)l_{\alpha }(k)\cos \;\theta _{\alpha ,\alpha +1}(k)%
\right] ^{\frac{1}{2}},  \notag
\end{gather}
where $L_{\alpha }^{+}(k)$ denotes the length of the half-edge of \ $%
|P_{T_{L}}|\rightarrow {M}$ issuing from the vertex $\varrho _{\alpha
}^{o}(k)$ towards the midpoint $W_{\alpha +1}(k)$.\ Similarly we can define
the length $L_{\alpha }^{-}(k)$, associated with the half-edge issuing from $%
W_{\alpha +1}(k)$ towards the vertex $\varrho _{\alpha +1}^{o}(k)$, \emph{%
viz.}, 
\begin{gather}
L_{\alpha }^{-}(k)\equiv \left| \overrightarrow{W_{\alpha +1}(k)\varrho
_{\alpha +1}^{o}(k)}\right|  \label{minushalf} \\
=\left[ \frac{1}{36}l_{\alpha +1}^{2}(k)+\frac{1}{9}l_{\alpha +2}^{2}(k)-%
\frac{1}{9}l_{\alpha +2}(k)l_{\alpha +1}(k)\cos \;\theta _{\alpha +1,\alpha
+2}(k)\right] ^{\frac{1}{2}},  \notag
\end{gather}
(note that $L_{\alpha }^{-}(k)$ for $\alpha =q(k)$ is given by $\left| 
\overrightarrow{\varrho _{1}^{o}(k)W_{1}(k)}\right| $). Finally, one can
also consider the half-edge of \ $|P_{T_{L}}|\rightarrow {M}$ issuing from
the midpoint $W_{\alpha ,\alpha +1}(k)$ of the edge $V_{\alpha }(k)V_{\alpha
+1}(k)$ towards the vertex $\varrho _{\alpha }^{o}(k)$. Its length is
provided by 
\begin{gather}
L_{\alpha ,\alpha +1}^{-}(k)\equiv \left| \overrightarrow{W_{\alpha ,\alpha
+1}(k)\varrho _{\alpha +1}^{o}(k)}\right| =\frac{1}{2}\left| \overrightarrow{%
O\varrho _{\alpha }^{o}(k)}\right| \\
=\frac{1}{6}\left[ l_{\alpha +1}^{2}(k)+l_{\alpha }^{2}(k)+2l_{\alpha
+1}(k)l_{\alpha }(k)\cos \;\theta _{\alpha ,\alpha +1}(k)\right] ^{\frac{1}{2%
}},  \notag
\end{gather}
where we have exploited the fact that the medians of the triangles $\Delta
_{\alpha +1}(k)$ are divided in the ratio $2:1$ by the barycenters $\varrho
_{\alpha }^{o}(k)$. From the relation 
\begin{gather}
l_{\alpha ,\alpha +1}^{2}(k)\equiv \left| \overrightarrow{V_{\alpha
}(k)V_{\alpha +1}(k)}\right| ^{2} \\
=l_{\alpha +1}^{2}(k)+l_{\alpha }^{2}(k)-2l_{\alpha +1}(k)l_{\alpha }(k)\cos
\;\theta _{\alpha ,\alpha +1}(k),  \notag
\end{gather}
(and similarly for $l_{\alpha +1,\alpha +2}^{2}(k)$), it follows that we can
conveniently rewrite $L_{\alpha }^{+}(k)$, $L_{\alpha }^{-}(k)$, and $%
L_{\alpha ,\alpha +1}^{-}(k)$ as 
\begin{equation}
36[L_{\alpha }^{+}(k)]^{2}=2l_{\alpha }^{2}(k)+2l_{\alpha ,\alpha
+1}^{2}(k)-l_{\alpha +1}^{2}(k),  \label{dual1}
\end{equation}

\begin{equation}
36[L_{\alpha }^{-}(k)]^{2}=2l_{\alpha +2}^{2}(k)+2l_{\alpha +1,\alpha
+2}^{2}(k)-l_{\alpha +1}^{2}(k),  \label{dual2}
\end{equation}

\begin{equation}
36[L_{\alpha ,\alpha +1}^{-}(k)]^{2}=2l_{\alpha }^{2}(k)+2l_{\alpha
+1}^{2}(k)-l_{\alpha ,\alpha +1}^{2}(k).  \label{dual3}
\end{equation}
Note that the quantities $(L_{\alpha -1}^{-})^{2}$ and $(L_{\alpha
}^{+})^{2} $ are not independent, being related by 
\begin{equation}
\sum_{\alpha =1}^{q(k)}[(L_{\alpha -1}^{-})^{2}-(L_{\alpha }^{+})^{2}]\equiv
0,
\end{equation}
(again a consequence of the geometry of the medians). Since the edge
connecting the barycenter $\varrho _{\alpha }^{o}(k)$ of the triangle $%
\Delta _{\alpha }(k)$ with the barycenter $\varrho _{\alpha +1}^{o}(k)$ of
the adjacent triangle $\Delta _{\alpha +1}(k)\ \ $must pass through the
midpoint $W_{\alpha +1}(k)$ of the edge $OV_{\alpha +1}(k)$, the length of
the edge $\varrho _{\alpha }^{o}(k)\varrho _{\alpha +1}^{o}(k)\in
|P_{T_{L}}|\rightarrow {M}$ is provided by 
\begin{equation}
L_{\alpha }(k)=L_{\alpha }^{-}(k)+L_{\alpha }^{+}(k),  \label{Lpiu}
\end{equation}
namely, 
\begin{gather}
L_{\alpha }(k)=\frac{1}{3\sqrt{2}}\left[ l_{\alpha }^{2}(k)+l_{\alpha
,\alpha +1}^{2}(k)-\frac{1}{2}l_{\alpha +1}^{2}(k)\right] ^{\frac{1}{2}}
\label{elletot} \\
+\frac{1}{3\sqrt{2}}\left[ l_{\alpha +2}^{2}(k)+l_{\alpha +1,\alpha
+2}^{2}(k)-\frac{1}{2}l_{\alpha +1}^{2}(k)\right] ^{\frac{1}{2}},  \notag
\end{gather}
The set of relations (\ref{dual1}), (\ref{dual2}), and (\ref{dual3}), (as
the index $k$ varies over all vertices of $|T_{l=a}|\rightarrow M$, or over
all $2$-cells of $|P_{T_{L}}|\rightarrow M$ ), allow a complete
transcription between the metric geometry of a Regge surface $%
|T_{l}|\rightarrow M$ and the geometry of the corresponding dual polytope $%
|P_{T_{L}}|\rightarrow M$.

For a $2$-dimensional dynamical triangulation $|T_{l=a}|\rightarrow M$ the
above analysis is only apparently trivial, since it allows us to discuss
what happens to the triple of half-edges $L_{\alpha -1}^{-}(k)$, $L_{\alpha
}^{+}(k)$, and $L_{\alpha ,\alpha +1}^{-}(k)$ when we deform the
edge-lengths of the generic equilateral triangle $\Delta _{\alpha }(k)\in
|T_{l=a}|\rightarrow M$. In particular, by differentiating \ (\ref{dual1}), (%
\ref{dual2}), and (\ref{dual3})\ with respect to the Regge edge-lengths $%
\{l_{\alpha }(k),l_{\alpha ,\alpha +1}(k),l_{\alpha +1}(k)\}$\ \ \ we get 
\begin{equation}
\left. dL_{\alpha }^{+}(k)\right| _{l=a}=\frac{1}{3\sqrt{3}}\left[
dl_{\alpha }(k)+dl_{\alpha ,\alpha +1}(k)-\frac{1}{2}dl_{\alpha +1}(k)\right]
,  \label{lin1}
\end{equation}

\begin{equation}
\left. dL_{\alpha -1}^{-}(k)\right| _{l=a}=\frac{1}{3\sqrt{3}}\left[
dl_{\alpha +1}(k)+dl_{\alpha ,\alpha +1}(k)-\frac{1}{2}dl_{\alpha }(k)\right]
,  \label{lin2}
\end{equation}
\begin{equation}
\left. dL_{\alpha ,\alpha +1}^{-}(k)\right| _{l=a}=\frac{1}{3\sqrt{3}}\left[
dl_{\alpha }(k)+dl_{\alpha +1}(k)-\frac{1}{2}dl_{\alpha ,\alpha +1}(k)\right]
.  \label{lin3}
\end{equation}
These relations are invertible and show that the information contained in
the deformation of the barycentrically dual polytope $|P_{T_{a}}|\rightarrow
M$ \ is the same as the information contained in the deformation of the
dynamical triangulation $|T_{l=a}|\rightarrow M$. Thus, if we denote by 
\begin{gather}
\frak{P}:\left\{ |T_{l}|\rightarrow M\right\} \rightarrow \left\{
|P_{T_{L}}|\rightarrow M\right\} , \\
\{l_{(j)}\}_{j=1}^{N_{1}(T)}\longmapsto \{L_{(j)}^{\pm }\}_{j=1}^{N_{1}(P)},
\notag
\end{gather}
the map which to the set of edge-lengths $\{l_{(j)}\}_{j=1}^{N_{1}(T)}$ of a
Regge surface associates the edge-lengths of the corresponding half-edges in
the barycentrically dual polytope, then from the invertibility of (\ref{lin1}%
), (\ref{lin2}), and (\ref{lin3})\ we immediately get the following

\begin{lemma}
The map $\frak{P}$ \ is a local isomorphism around any given dynamical
triangulation $|T_{l=a}|\rightarrow M$. \ In other words, \ there is a
one-to-one correspondence between the local deformations of an equilateral
polytope $|P_{T_{a}}|\rightarrow M$ and the Regge triangulations that can be
obtained through the local deformations of the edge-lengths of the dynamical
triangulation $|T_{l=a}|\rightarrow M$ associated with $|P_{T_{a}}|%
\rightarrow M$.
\end{lemma}

Later on we shall discuss in greater detail the geometry of such
deformations around a given $|P_{T_{a}}|\rightarrow M$, and show that is
strictly related with the Witten-Kontsevich model.

\subsection{Quadratic differentials and singular Euclidean structures}

In order to to provide a better visualization of the correspondence between
singular Euclidean structures, dynamical triangulations, Regge surfaces and
moduli spaces of punctured Riemann surfaces we need several basic
definitions from surface theory. We apologize to the expert reader who may
skip a large part of this section.

We start by recalling that \ a holomorphic quadratic differential, $\psi $,
(a transverse traceless rank two tensor), on a Riemann surface $M$ is
defined, in a locally uniformizing complex coordinate chart $(U,\zeta )$, by
a holomorphic function $\mu :U\rightarrow \mathbb{C}$ such that $\psi =\mu
(\zeta )d\zeta \otimes d\zeta $. A quadratic differential $\psi $\ is said
to have order $m$ at a point $p\in M$ if, \ in a neighborhood of $p$, we can
write $\psi =\mu (\zeta )d\zeta \otimes d\zeta $, with $\mu (\zeta )$
possessing a zero of order $m$ at $p$. Note that, on a surface of genus $g>0$%
, the number of zeros of a non-trivial quadratic differential $\psi $ is
given by $4g-4$ (counting multiplicities). For genus $g>0$ the complex
vector space of quadratic differentials, $Q(M)$, is non-empty with complex
dimension $\dim _{\mathbb{C}}\,Q(M)=3g-3$, ($\dim _{\mathbb{C}}\,Q(M)=1$,
for $g=1$). The geometry of $Q(M)$ is directly related with the
characterization of the Teichm\"{u}ller space of $M$, $\frak{T}_{g}(M)$, the
space of all conformal structures on $M$ under the equivalence relation
given by pullback by diffeomorphisms isotopic to the identity map $%
id:M\rightarrow M$. It is well known that $\frak{T}_{g}(M)$ is a smooth
finite dimensional manifold that can be identified with a $6g-6$ \ ($\mathbb{%
R}$)-dimensional cell defined by the open unit ball (in a suitable norm) in
the space of quadratic differentials. Conversely, the tangent space to $%
\frak{T}_{g}(M)$ at a reference quadratic differential $\psi $, is $\mathbb{C%
}$-anti-linear isomorphic to$\ Q(M)$, in other words $Q(M)$ can be
canonically identified with the cotangent space to $\frak{T}_{g}(M)$. If $%
\pi _{1}(M)$ denotes the fundamental group of the surface $M$, then\ the
quotient of $\frak{T}_{g}(M)$ by the action of the outer automorphism group
of $\pi _{1}(M)$, \emph{i.e.}, 
\begin{equation}
\frak{M}_{g}=\frak{T}_{g}(M)/Out(\pi _{1}(M))
\end{equation}
is the Riemann moduli space parametrizing conformal equivalence classes of
Riemann surfaces of genus $g$. If we fix $\lambda $ distinct points $%
x_{1},...,x_{\lambda }\in M$, corresponding to which $M$ is punctured,\ then
the corresponding moduli space acquires one extra (complex) dimension for
each puncture, i.e., 
\begin{equation}
\dim _{\mathbb{C}}\,\frak{M}_{g},_{\lambda }=3g-3+\lambda .
\end{equation}
The Deligne-Mumford stable curve compactification of such a moduli space is
obtained by adding to $\frak{M}_{g},_{\lambda }$ a suitable set of singular
noded surfaces according to a prescription that makes $\frak{M}%
_{g},_{\lambda }$ into a compact (orbifold) space $\overline{\frak{M}}%
_{g},_{\lambda }$. The stability condition guarantees that the surface has
only a finite automorphism group, moreover an important feature of \ $%
\overline{\frak{M}}_{g},_{\lambda }$ is that it compactifies $\frak{M}%
_{g},_{\lambda }$ without allowing the marked points to come together.
Roughly speaking, when points on a (smooth) surface approach each other, the
surface sprouts off one or more components and the points distribute
themselves on these new components. Recall that on $\frak{M}_{g},_{\lambda }$
one naturally introduces a set of cohomology classes of degree two, $c_{1}(%
\mathcal{L}_{i})$, $i=1,...,\lambda $, defined by the Chern class of the
line bundle $\mathcal{L}_{i}$ over $\frak{M}_{g},_{\lambda }$ whose fiber at 
$[M,x_{1},...,x_{\lambda }]\in \frak{M}_{g},_{\lambda }$ is the cotangent
space to $M$ at $x_{i}$. In terms of such cohomology classes one defines
[5],[6] intersection numbers $\left\langle \tau _{\delta _{1}}...\tau
_{\delta _{\lambda }}\right\rangle $ over $\overline{\frak{M}}_{g},_{\lambda
}$ by setting 
\begin{equation}
\left\langle \tau _{\delta _{1}}...\tau _{\delta _{\lambda }}\right\rangle
\doteq \int_{\overline{\frak{M}_{g},_{\lambda }}}c_{1}(\mathcal{L}%
_{1})^{\delta _{1}}\wedge ...\wedge c_{1}(\mathcal{L}_{\lambda })^{\delta
_{\lambda }},\;  \label{Kontnumb}
\end{equation}
where the integral is zero unless the sequence of non-negative integers $%
\{\delta _{i}\}_{i=1}^{\lambda }$\ is such that $\sum_{k=1}^{\lambda }\delta
_{k}=\lambda +3g-3$.\ Since $\overline{\frak{M}}_{g},_{\lambda }$ is an
orbifold, the $\left\langle \tau _{\delta _{1}}...\tau _{\delta _{\lambda
}}\right\rangle $ are positive rational numbers.\ \ Their properties as well
as the geometry itself of \ $\overline{\frak{M}}_{g},_{\lambda }$ are
strictly connected to a combinatorial stratification of $\overline{\frak{M}}%
_{g},_{\lambda }$ in terms of the graphical data describing inequivalent
quadratic differentials. The rationale underlying such stratification is the
observation that a quadratic differential $\psi $ may be pictured by a
transverse measured foliation which induces on $M$ singular Euclidean
structures whose moduli can be connected with $\overline{\frak{M}}%
_{g},_{\lambda }$. For a better visualization of such concepts, recall that
a measured foliation with singularities $x_{1},...,x_{k}$, respectively of
order $m_{1},...,m_{k}$, \ is a one-dimensional foliation of $%
M-\{x_{1},...,x_{k}\}$ with $\{m_{j}+2\}$ pronged singularities at $%
\{x_{j}\} $, together with a transverse measure which assigns to each arc
transverse to the foliation a nonnegative real number, such that the natural
maps between transversals are measure-preserving. The foliation by
horizontal lines in $\mathbb{C}$ with transverse measure $|dy|$, and more
generally\ the horizontal and vertical trajectories of a holomorphic
quadratic differentials provide the typical example of measured foliations.
Explicitly, away from the discrete set of the zeros of \ $\psi \in
Q(M)-\{0\} $ we can locally choose a canonical conformal coordinate $\zeta $
(unique up to $\zeta \mapsto \pm \zeta +const$) by integrating the
holomorphic $1$-form $\sqrt{\psi }$, \emph{i.e}., \ 
\begin{equation}
\zeta (z)=\int^{z}\sqrt{\psi },
\end{equation}
so that $\psi =d\zeta \otimes d\zeta $. In such a chart, the measured
foliation associated with $\psi $ becomes the usual one whose leaves are the
horizontal lines $(\{\func{Im}(\zeta )=const\},|d\func{Im}(\zeta )|)$.\
Similarly, the local sets $\{\func{Re}(\zeta )=const\}$ endowed with the
measure $|d\func{Re}(\zeta )|$ piece together to form \ the vertical
measured foliation associated with $\psi $. At the generic zero $\sigma _{j%
\text{ }}$of $\psi $ of order $m_{j}$ we get a cyclically ordered set of $%
m_{j}+2$ pronged singularities counting the number of vertical and
horizontal leaves emanating from the singularity. The cyclic order of the
prongs (or half-edges) at the singularities is determined by the orientation
of the Riemann surface $M$.

If the measured foliation generated by a quadratic differential $\psi \in
Q(M)-\{0\}$ has closed horizontal leaves (up to a set of measure zero on the
surface), then such a $\psi $ decomposes the surface $M$ into the maximal
ring domains foliated by the closed leaves, (typically annuli or punctured
disks). As is well known, such a propery characterizes the Jenkins-Strebel
(JS) quadratic differentials. It is important to realize that if we
introduce the local parametrization $\zeta (z)$, the measure associated with 
$|\psi |$ naturally induces on $M$ a structure of singular flat Riemann
surface. Explicitly, let $x_{1},...,x_{k}$ denote the zeros of $\psi $ with
respective multiplicities $m_{1},...,m_{k}$, then the divisor 
\begin{equation}
\sum_{j=1}^{k}\left( \frac{\theta (j)}{2\pi }-1\right) \,x_{j},
\label{visor}
\end{equation}
with $\theta (j)\doteq (m(j)+2)\pi ,$ defines on $M$ a singular Euclidean
structure with conical singularities $\theta (j)$ supported on the set of $k$
zeros, $x_{1},...,x_{k}$, of $\psi $\ . The corresponding metric is provided
by 
\begin{equation}
\mathbf{ds}^{2}=\left| \psi \right| =\left( \prod_{j=1}^{k}\left| \zeta
-\left( \zeta (x_{j})\right) \right| ^{m_{j}}\right) \left| d\zeta \right|
^{2},
\end{equation}
which for the differential $dz^{2}$ is just the Euclidean metric. \ Note
that the divisor (\ref{visor}) has degree given by 
\begin{equation}
\sum_{j=1}^{k}\left( \frac{(m(j)+2)\pi }{2\pi }-1\right) =\sum_{j=1}^{k}%
\frac{m(j)}{2}=-\chi (M),
\end{equation}
as expected. The special feature of the singular Euclidean structure
associated with a JS quadratic differential $\left| \psi \right| $ \ is that
it can be naturally put in correspondence to to the cell decomposition of
the surface $M$ associated with the rigid polytopes $|P_{T_{a}}|\rightarrow
M $. This is a well-known result that can be considered as a particular case
of the Troyanov-Picard theorem, and which in the DT framework can be
explicitly described according to

\begin{proposition}
Let \ $|T_{l=a}|\rightarrow M$ be a dynamically triangulated surface of
genus $g$, with a given set of ordered curvature assignments $\{q(i),\sigma
^{0}(i)\}_{i=1}^{N_{0}(T)}$ over its $N_{0}(T)$ labelled vertices. Let $%
|P_{T_{a}}|\rightarrow M$ the equilateral polygonalization of $\ $the
surface $\ M$ associated to the dual polytope of $|T_{l=a}|\rightarrow M$.
Identify the labelled vertices \ $\{\sigma ^{0}(i)\}_{i=1}^{N_{0}(T)}$\ \ of
\ \ $|T_{l=a}|\rightarrow M$ with a corresponding \ set of marked points of $%
\ M$. \ If 
\begin{equation}
\left\{ 
\begin{tabular}{l}
$g\geq 0$ \\ 
$N_{0}(T)\geq 1$ \\ 
$2-2g-N_{0}(T)<0$%
\end{tabular}
\right. ,
\end{equation}
then there is a unique JS quadratic differential $\psi \in Q(M)-\{0\}$ on $M$
satisfying the following conditions: (i) $\psi $ is holomorphic on $%
M\backslash \{\sigma ^{0}(i)\}_{i=1}^{N_{0}(T)}$. (ii) $\psi $ has $N_{2}(T)$
zeros (of order $1$). (iii) $\psi $ has a double pole at each $\sigma
^{0}(k)\in \{\sigma ^{0}(i)\}_{i=1}^{N_{0}(T)}$. (iv) The union of all
noncompact horizontal leaves of $\psi $ generate the $1$-skeleton of $%
|P_{T_{a}}|\rightarrow M$. (v) Every compact horizontal leaf of $\psi $ is a
simple loop $\frak{L}(\sigma ^{0}(k))$ circling around a vertex $\sigma
^{0}(k)\in \{\sigma ^{0}(i)\}_{i=1}^{N_{0}(T)}$, and its (constant) length
is proportional to the corresponding curvature assignment $q(k)$, i.e., 
\begin{equation}
\left( \frac{\sqrt{3}}{3}a\right) q(k)=\oint_{L(\sigma ^{0}(k))}\sqrt{\psi }.
\label{loop}
\end{equation}
\ Note in particular that the simple loop $\frak{L}(\sigma ^{0}(k))$ which
hits the corresponding set of zeros of \ $\psi $ is the boundary of the $%
q(k) $-gonal $2$-cell $\in |P_{T_{a}}|\rightarrow M$ baricentrically dual to
the vertex $\sigma ^{0}(k)\in |T_{l=a}|\rightarrow M$.
\end{proposition}

\emph{Proof}. This is basically a rephrasing, in the framework of dynamical
triangulation theory, of the classical result of Strebel [7], (we again
warmly recommend the remarkable paper [5] for a detailed exposition of
Strebel theory, ribbon graph theory, moduli space, and all that). Note that,
according to (\ref{loop}), under the correspondence 
\begin{equation}
(|P_{T_{a}}|\rightarrow M)\rightarrow (M,\psi ),
\end{equation}
the $q(k)$-gonal $2$-cell $\in |P_{T_{a}}|\rightarrow M$ dual to the vertex $%
\sigma ^{0}(k)\in |T_{l=a}|\rightarrow M$ \ goes into an infinite tube whose
constant transverse length is $(\frac{\sqrt{3}}{3}a)q(k)$. In other words,
each dual $2$-cell of the dynamical triangulation $|T_{l=a}|\rightarrow M$
gives rise to a punctured disk in the corresponding Riemann surface $(M,\psi
)$. Thus the given cut-off $a$\ associated with the fixed edge-length of the 
$N_{2}(T)$ triangles of a dynamical triangulation, is represented in the
corresponding Riemann surface $(M,\psi )$\ by the presence of $N_{0}(T)$
punctured disks foliated by compact horizontal leaves $\simeq S^{1}$ each
one of \emph{constant} length proportional to $a$. In other words, the tubes
of constant section $\propto a$ are the counterpart, in surface theory, of
the $Diff$-invariant cut-off characterizing dynamical triangulations. $%
\blacksquare $

\bigskip

As is well known, from a mathematical point of view the theory of JS
quadratic differentials allows the parametrization of the holomorphic
structure of a Riemann surface into the combinatorial data of metric ribbon
graphs. Such a parametrization defines a bijective mapping (a homeomorphism
of orbifolds) between the space of ribbon graphs $RGB_{g,N}^{met}$ and the
moduli space $\frak{M}_{g},_{N}$ of Riemann surfaces $M$ of genus $g$ with $%
N $ ordered marked points (punctures) [1], [5], 
\begin{eqnarray}
\frak{M}_{g},_{N}\times \mathbb{R}_{+}^{N} &\longleftrightarrow
&RGB_{g,N}^{met} \\
(M,L_{i}) &\longleftrightarrow &\Gamma  \notag
\end{eqnarray}
where $(L_{1},...,L_{N})$ is an ordered n-tuple of positive real numbers and 
$\Gamma $ is a metric ribbon graphs with $N$ labelled boundary lengths $%
\{L_{i}\}$ defined by the corresponding JS quadratic differential. In
particular, we have

\begin{proposition}
Let 
\begin{equation}
\mathcal{DT}[\{q(k)\}_{k=1}^{N_{0}}]\doteq \left\{ |T_{l=a}|\rightarrow
M\;:q(\sigma ^{0}(k))=q(k),\;k=1,...,N_{0}(T)\right\} .
\end{equation}
\ \ denote the set of \ distinct dynamically triangulated surfaces of genus $%
g$, with a given set of ordered curvature assignments $\{q(i),\sigma
^{0}(i)\}_{i=1}^{N_{0}(T)}$ over its $N_{0}(T)$ labelled vertices.\ If \ $%
g\geq 0$, $2-2g-N_{0}(T)>0$, then\ there is an injective mapping 
\begin{equation}
\mathcal{DT}[\{q(k)\}_{k=1}^{N_{0}}]\longrightarrow \frak{M}%
_{g},_{N_{0}}\times \left( \frac{\sqrt{3}}{3}a\right) \mathbb{N}_{+}^{N_{0}}
\label{injection}
\end{equation}
which is defined by associating to the JS quadratic differential $\psi $,
defined by the dual polygonalization of a $|T_{l=a}|\rightarrow M\in 
\mathcal{DT}[\{q(k)\}_{k=1}^{N_{0}}]$, \ the corresponding punctured Riemann
surface $(M/\{\sigma ^{0}(i)\}_{i=1}^{N_{0}(T)}\},\psi )$.
\end{proposition}

\emph{Proof}. The injectivity of the map is obvious from the unicity of the
JS differential associated with the dual polytope corresponding to $%
|T_{l=a}|\rightarrow M\in \mathcal{DT}[\{q(k)\}_{k=1}^{N_{0}}]$.
Surjectivity fails since the metric ribbon graphs associated with
triangulations in $\mathcal{DT}[\{q(k)\}_{k=1}^{N_{0}}]$ do not span the
whole $RGB_{g,N_{0}}^{met}$, but only a subset of \ $RGB_{g,N_{0}}^{met}$
generated by ribbon graphs whose labelled boundaries have lengths which are
integer multiples of \ $(\frac{\sqrt{3}}{3}a)$. $\blacksquare $

\bigskip

In this connection, it is perhaps worthwhile noticing that the image of the
map (\ref{injection}) coincides with the set of algebraic curves defined
over the algebraic closure $\overline{\mathbb{Q}}$ of the field of rational
numbers, (\emph{i.e.}, over the set of complex numbers which are roots of
non-zero polynomials with rational coefficients). This latter result, due to
V. A. Voevodskii and G. B. Shabat [8], establishes a remarkable bijection
between dynamical triangulations and curves over algebraic number fields.
The proof in [8] exploits the characterization of the collection of
algebraic curves defined over $\overline{\mathbb{Q}}$ provided by Belyi's
theorem (see e.g., [5]), according to which a nonsingular Riemann surface $M$
has the structure of an algebraic curve defined over $\overline{\mathbb{Q}}$
if and only if there is a holomorphic map (a branched covering of $M$ over
the sphere) 
\begin{equation}
f:M\rightarrow \mathbb{CP}^{1}
\end{equation}
that is ramified only at $0,1$ and $\infty $, (such maps are known as Belyi
maps). It is worthwhile remarking that the inverse image of the line segment 
$[0,1]\subset \mathbb{CP}^{1}$ under a Belyi map is a Grothendieck's \emph{%
dessin d'enfant}, thus dynamically triangulated surfaces are eventually
connected with the theory of the Galois group $Gal(\overline{\mathbb{Q}}/%
\mathbb{Q})$ action on the branched coverings $f:M\rightarrow \mathbb{CP}%
^{1} $. The correspondence between Belyi maps, \emph{dessin d'enfant} and JS
quadratic differentials has been recently analized in depth by M. Mulase and
M. Penkava [5], an equally inspiring paper is [9] by M. Bauer and C.
Itzykson.

\bigskip

\subsection{Triangulations and Moduli spaces of polygons}

The relation between a dynamical triangulation and a punctured Riemann
surface $(M,\psi )$ appears quite rigid since it exploits both the
equilateral structure and the datum of the perimeters $(\frac{\sqrt{3}}{3}%
a)q(k)$\ of the polygons $\in |P_{T_{a}}|\rightarrow M$. If we vary the edge
lengths ($=(\frac{\sqrt{3}}{3}a)$) of the polygonal loops $\{\frak{L}(\sigma
^{0}(k))\}$ continuosly while leaving the given curvature assignments $%
\{q(i),\sigma ^{0}(i)\}_{i=1}^{N_{0}(T)}$fixed, how does this affect the
complex structure of \ the Riemann surface $(M,\psi )$? What is at issue
here is how the complex structure changes as we deform $|T_{l=a}|\rightarrow
M$ around the generic vertex $\sigma ^{0}(k)$ by keeping fixed the adjacency
but varying the edge-lengths of the $q(k)$ triangles incident on $\sigma
^{0}(k)$, i.e., by considering Regge triangulations associated with $%
|T_{l=a}|\rightarrow M$. The rationale of this question is to ensure that
one is looking at persistent rather than accidental features of \ the
relation between dynamical triangulations, Regge triangulations, and Riemann
surfaces. According to lemma 3 there is a one to one correspondence between
local deformations of $|T_{l=a}|\rightarrow M$ and local deformations of the
equilateral polytope $|P_{T_{a}}|\rightarrow M$. Thus in order to discuss
the above issue, it is sufficient to consider the deformational degrees of
freedom of the generic equilateral polygon $p_{eq}\left( k,a\right) \in
|P_{T_{a}}|\rightarrow M$. Let 
\begin{equation}
\mathcal{P}_{q(k)}\doteq \left\{ p(k):q(k)-gons\;in\;\mathbb{E}^{2}\right\}
\end{equation}
be the space\ of all (not necessarily equilateral) polygons $p\left(
k\right) $ with $q(k)$ labelled vertices $v_{\alpha }=X^{\alpha }(k)+\sqrt{-1%
}Y^{\alpha }(k)$, $v_{q(k)+1}=v_{1}$, in the Euclidean plane $\mathbb{E}^{2}$
($\simeq \mathbb{C}$). Note that as we circle around $p(k)$ in the
counterclockwise direction, such vertices are supposed to be ordered up to a
cyclic permutation. We shall consider also the space $[\mathcal{P}_{q(k)}]$
of equivalence classes of all polygons $p\left( k\right) \in \mathcal{P}%
_{q(k)}$ with any two polygons identified if there exists an orientation
preserving similarity of \ $\mathbb{E}^{2}$ \ which sends vertices of one
polygon to vertices of \ the other one. Let 
\begin{gather}
Z^{\alpha }(k)\doteq v_{\alpha +1}-v_{\alpha }=(X^{\alpha +1}(k)-X^{\alpha
}(k))+\sqrt{-1}(Y^{\alpha +1}(k)-Y^{\alpha }(k))\in \mathbb{C} \\
L_{\alpha }(k)=\sqrt{(X^{\alpha +1}(k)-X^{\alpha }(k))^{2}+(Y^{\alpha
+1}(k)-Y^{\alpha }(k))^{2}},  \notag
\end{gather}
($\alpha =1,...,q(k)$), respectively denote the $q(k)$ edges of the polygon $%
p(k)$ and their lengths. According to (\ref{elletot}) we can explicitly
write the $L_{\alpha }(k)$ in terms of the edge-lenghts $\{l_{\alpha }(k)\}$
of the Regge triangulation $|T_{l}|\rightarrow M$ which, upon barycentrical
dualization, provides $|P_{T_{L}}|\rightarrow M$.

If we allow all possible deformations of $p(k)$ except the collapse to a
point, a generic $q(k)$-gon $p(k)$ is described by the $q(k)-1$ complex
coordinates 
\begin{equation}
p(k)=(Z^{1}(k),...,Z^{q(k)-1}(k))\in \mathbb{C}^{q(k)-1}-\{0\},
\end{equation}
(since the closing condition $\sum_{\alpha =1}^{q(k)}Z^{\alpha }(k)=0$ for
the edges of the polygon $p(k)$ implies that $Z^{q(k)}(k)=-\sum_{\alpha
=1}^{q(k)-1}Z^{\alpha }(k)$). Moreover, any two polygons $p(k)$ and $%
\widetilde{p(k)}$ such that 
\begin{equation}
(\widetilde{Z}^{1}(k),...,\widetilde{Z}^{q(k)-1}(k))=\lambda
(Z^{1}(k),...,Z^{q(k)-1}(k)),  \label{proj}
\end{equation}
for some $\lambda \in \mathbb{C}-\{0\}$ define the same point in $[\mathcal{P%
}_{q(k)}]$. In other words, the orbit space $[\mathcal{P}_{q(k)}]$ is
canonically isomorphic to the complex projective space over the hyperplane $%
\sum_{\alpha =1}^{q(k)}Z^{\alpha }(k)=0\subset \mathbb{C}^{q(k)}$, namely to 
$\mathbb{CP}^{q(k)-2}$. From such an identification it follows that the
assignment of $\ (Z^{1}(k),...,Z^{q(k)-1}(k))$ to the equivalence class of
polygons $[p(k)]$ it defines in $\mathbb{CP}^{q(k)-2}$ is nothing but the
natural projection $\ \ \pi :\mathbb{C}^{q(k)-1}-\{0\}\rightarrow \mathbb{CP}%
^{q(k)-2}$, [10]. In particular a point in the inverse image 
\begin{equation}
\pi ^{-1}([p(k)])\simeq \mathbb{C}^{\ast }\doteq \mathbb{C}-\{0\},
\end{equation}
is the polygon locally described by 
\begin{equation}
p(k)=(\frac{Z^{1}}{Z^{\nu }},...,\widehat{\frac{Z^{\nu }}{Z^{\nu }}},...%
\frac{Z^{q(k)-1}}{Z^{\nu }};Z^{\nu })
\end{equation}
with $\{\frac{Z^{1}}{Z^{\nu }}\}$, $\nu \neq 1$, being the local coordinates
of an equivalence class of polygons, $\widehat{}$ denoting an omitted
coordinate, and $Z^{\nu }\in $ $\mathbb{C}^{\ast }$. Since any polygon $p(k)$
may occur as the the $q(k)$-gon dual to a vertex of a Regge triangulation,
we have the following obvious characterization

\begin{proposition}
The set of (labelled) $q(k)$-gons baricentrically dual to the $\sigma
^{0}(k) $ vertex of a Regge triangulation is described by the principal $%
\mathbb{C}^{\ast }$-bundle 
\begin{equation}
\mathcal{P}_{q(k)}:\mathbb{C}^{q(k)-1}-\{0\}\overset{\pi }{\longrightarrow }%
\mathbb{CP}^{q(k)-2}.
\end{equation}
\end{proposition}

Note that we can equivalently describe such a set of polygons by means of
the associated canonical line bundle over $\mathbb{CP}^{q(k)-2}$ (the dual
Hopf bundle).

From $[\mathcal{P}_{q(k)}]\simeq \mathbb{CP}^{q(k)-2}$ it also follows that $%
[\mathcal{P}_{q(k)}]$\ is a compact connected complex manifold \ that can be
equipped with the standard Fubini-Study K\"{a}hler form $\omega _{FS}(k)$, 
\begin{equation}
\left. \omega _{FS}(k)\right| _{U_{q(k)-1}}=\frac{\sqrt{-1}}{2}\left\{ \frac{%
dZ^{\alpha }(k)\wedge dZ\overline{^{\alpha }}(k)}{Z^{\beta }(k)Z\overline{%
^{\beta }}(k)}-\frac{Z\overline{^{\alpha }}(k)Z^{\beta }(k)dZ^{\alpha
}(k)\wedge dZ\overline{^{\beta }}(k)}{(Z^{\mu }(k)Z\overline{^{\mu }}(k))^{2}%
}\right\} .  \label{FS}
\end{equation}
Such remark implies that [10], if $[Z^{1}(k),...,Z^{q(k)-1}(k)]$ denote the
homogeneous coordinates representing the equivalence class of polygons $%
[p(k)]$, then each polygon representing $[p(k)]\in \lbrack \mathcal{P}%
_{q(k)}]$ can be continuously deformed to another polygon, representing a
distinct equivalence class $[\widehat{p(k)]}\in \lbrack \mathcal{P}_{q(k)}]$%
, by means of an interpolating curve of poligons 
\begin{gather}
\mathbb{R}\supset I\longrightarrow \lbrack \mathcal{P}_{q(k)}] \\
t\mapsto \lbrack Z^{1}(k,t),...,Z^{q(k)-1}(k,t)]=[p(k;t)]\in \lbrack 
\mathcal{P}_{q(k)}],  \notag
\end{gather}
which can be chosen to be a geodesic with respect to the Fubini-Study metric
associated with $\omega _{FS}(k)$. Note that [10], since $[\mathcal{P}%
_{q(k)}]$ is compact, such a geodesic segment has length bounded above by a
constant which does not depend from $[$ $p(k)]$ and $[\widehat{p(k)]}$. \
Since (\ref{FS}) is invariant under (\ref{proj}), there is also a natural
action of $\mathbb{R}^{+}$ on $[\mathcal{P}_{q(k)}]$\ which corresponds to
an overall rescaling of \ the edge-lengths $(L_{1}(k),...,L_{q(k)}(k))\in 
\mathbb{R}^{q(k)}$ \ by a common factor $\eta =|\lambda |\in \mathbb{R}^{+}$%
. Thus, without loss of generality,\ we can require that the perimeter of \
the $q(k)$-gon \ $p(k,t)$ remains fixed (and equal to the perimeter of $%
p_{eq}(k,a)$)\ while deforming its edges,\emph{\ i.e.}, 
\begin{equation}
\sum_{\alpha =1}^{q(k)}L_{\alpha }(k,t)=(\frac{\sqrt{3}}{3}a)q(k),\;\forall
t\in I\subset \mathbb{R}.  \label{per}
\end{equation}
Moreover, any such a deformation is defined up to the action of a cyclic
permutation \ $\alpha _{1},...,\alpha _{q(i)}=s(1,...,q(k))$, \emph{i.e.}, 
\begin{equation}
(L_{1}(k),...,L_{q(k)}(k))\simeq (L_{\alpha _{1}}(k),...,L_{\alpha
_{q(k)}}(k)),
\end{equation}
where $s$ is an element of the symmetric group $\frak{S}_{q(i)}$ over $q(i)$
elements. This is a convenient moment for discussing more in detail the map 
\begin{gather}
\Im _{k}:\mathcal{P}_{q(k)}\longrightarrow \mathbb{R}^{q(k)}  \label{polmap}
\\
(Z^{\alpha }(k))\longmapsto \Im _{k}(Z^{\alpha }(k))=\left(
L_{1}(k),...,L_{q(k)}(k)\right)  \notag
\end{gather}
which assigns to each polygon $p(k)$, in the principal bundle $\mathcal{P}%
_{q(k)}$, the $q(k)$-tuple of its edge lengths $\{L_{\alpha }(k)\}$. The
image of $\Im _{k}$ is contained in the domain $D_{q(k)}\subset $ $\mathbb{R}%
^{q(k)}$ defined by 
\begin{gather}
\sum_{\alpha =1}^{q(k)}L_{\alpha }(k,t)=(\frac{\sqrt{3}}{3}a)q(k),
\label{lati} \\
0<L_{\alpha }(k,t)\leq (\frac{\sqrt{3}}{6}a)q(k),  \notag
\end{gather}
$\forall t\in I\subset \mathbb{R}$, (the upper bound on the $L_{\alpha }(k)$
follows from the triangle inequalities). The polygon $p_{eq}(k,a)$ dual to $%
\sigma ^{0}(k)\in |T_{l=a}|\rightarrow M$ belongs to the set of polygons
defined by the pre-image $\Im _{k}^{-1}((\frac{\sqrt{3}}{3}a),...,(\frac{%
\sqrt{3}}{3}a))\subset \mathbb{C}^{q(k)-1}-\{0\}$. However, note that $\Im
_{k}^{-1}((\frac{\sqrt{3}}{3}a),...,(\frac{\sqrt{3}}{3}a))$ may contain also
degenerate polygons (all with the same given perimeter) defined by those $%
q(k)-$gons which collapse into a straight line, (\emph{e.g.}, a
parallelogram with equilateral sides of length $(\frac{\sqrt{3}}{3}a)$ \ may
smoothly \emph{collapse}\ into a segment of length $2(\frac{\sqrt{3}}{3}a)$%
). It is worth noticing that [10] the set of degenerate polygons appears in $%
\mathcal{P}_{q(k)}$ as the fixed points set of the involution 
\begin{gather}
\mathcal{P}_{q(k)}\rightarrow \mathcal{P}_{q(k)}  \label{cconj} \\
(\{Z^{\alpha }(k)\}_{k=1}^{N_{0}})\rightarrow (\{Z\overline{^{\alpha }}%
(k)\}_{k=1}^{N_{0}}),  \notag
\end{gather}
induced by the complex conjugation. An elementary application of the results
of M. Kapovich amd J. Millson [10] (pp. 138), provides a suitable
characterization of the set of polygons in $\Im _{k}^{-1}(\{L_{\alpha
}(k,t)\})$ in a neighborhood of the equilateral $q(k)-$gon $p_{eq}(k,a)$
with edge-lengths $\{L_{\alpha }(k,t)\}=\{(\frac{\sqrt{3}}{3}a)\}$.

\begin{proposition}
In the principal bundle $\mathcal{P}_{q(k)}$\ there exists a neighborhood $%
U_{k}$ of the equilateral $q(k)-$gon \ $[p_{eq}(k,a)]$\ such that 
\begin{equation}
\Im _{k}^{-1}(\{L_{\alpha }(k,t)\})\cap U_{k},
\end{equation}
is a smooth submanifold of $\mathcal{P}_{q(k)}$\ \ \ \ whose tangent space
at $p_{eq}(k,a)$ is characterized by 
\begin{gather}
\left. {\Large T}\left[ \Im _{k}^{-1}(\{L_{\alpha }(k,t)\})\cap U_{k}\right]
\right| _{p_{eq}(k,a)}= \\
\left\{ (\xi ^{1},...,\xi ^{q(k)-1})\in \mathbb{C}^{q(k)-1}:\left. \func{Re}[%
Z^{\alpha }\xi \overline{^{\alpha }}]\right| _{\alpha
=1,...,q(k)-1}=0\,\right\} .  \notag
\end{gather}
\end{proposition}

In other words, $\Im _{k}:\mathbb{C}^{q(k)-1}-\{0\}\longrightarrow \mathbb{R}%
^{q(k)}$ is a smooth submersion in a neighborhood of $p_{eq}(k,a)$, and $%
(\Im _{k}^{-1}(\{L_{\alpha }(k,t)\}))\cap U_{k}$ is a smooth submanifold of
\ $\mathcal{P}_{q(k)}\mathbb{\ }$around \ $[p_{eq}(k,a)]$.\ In order to
prove such result, let us define a deformed $q(k)-$gon, in a neighborhood of
the equilateral polygon $p_{eq}(k,a)$, by smoothly varying its first $q(k)-1$
sides, while keeping its perimeter fixed, \emph{i.e.}, 
\begin{gather}
Z^{\alpha }(t)\doteq \left. Z^{\alpha }+t\xi ^{\alpha }\right| _{\alpha
=1,...,q(k)-1}, \\
Z^{q(k)}(t)\doteq -\sum_{\alpha =1}^{q(k)-1}Z^{\alpha }(t)  \notag
\end{gather}
with $|Z^{\alpha }|=(\frac{\sqrt{3}}{3}a)$, and $\ \sum_{\nu
=1}^{q(k)}Z^{\nu }(t)=(\frac{\sqrt{3}}{3}a)q(k)$, $\ \forall \ t\geq 0$.
Thus 
\begin{equation}
\Im _{k}(Z^{\alpha }(t))=(|Z^{1}(t)|,...,|Z^{q(k)-1}(t)|,(\frac{\sqrt{3}}{3}%
a)q(k)-\sum_{\alpha =1}^{q(k)-1}|Z^{q(k)}(t)|),
\end{equation}
and the corresponding tangent mapping is provided by 
\begin{gather}
D\Im _{k}\cdot \xi ^{\alpha }\doteq \left( \{\frac{d}{dt}|Z^{\alpha
}(t)|_{t=0}\}_{\alpha =1}^{q(k)-1},\frac{d}{dt}|Z^{q(k)}(t)|_{t=0}\right) \\
=\left( \left\{ \frac{\func{Re}[Z^{\alpha }\xi \overline{^{\alpha }}]}{%
|Z^{\alpha }|}\right\} _{\alpha =1}^{q(k)-1},-\sum_{\alpha =1}^{q(k)-1}\frac{%
\func{Re}[Z^{\alpha }\xi \overline{^{\alpha }}]}{|Z^{\alpha }|}\right) . 
\notag
\end{gather}
Since $p_{eq}(k,a)$ is non-degenerate, $D\Im _{k}\cdot \xi ^{\alpha }$ is
surjective near $p_{eq}(k,a)$. This proves that \ $\Im _{k}^{-1}(\{L_{\alpha
}(k,t)\})$ is a smooth submanifold of $\mathbb{C}^{q(k)-1}-\{0\}$\ with
tangent space provided by 
\begin{gather}
\ker D\Im _{k}\cdot \xi ^{\alpha }=  \label{ker} \\
\left\{ (\xi ^{1},...,\xi ^{q(k)-1})\in \mathbb{C}^{q(k)-1}:\left. \func{Re}[%
Z^{\alpha }\xi \overline{^{\alpha }}]\right| _{\alpha
=1,...,q(k)-1}=0\,\right\} .  \notag
\end{gather}
Note that $\dim _{\mathbb{R}}\,\ker (D\Im _{k}\cdot \xi ^{\alpha })=q(k)-1$
and $\dim _{\mathbb{R}}\,\func{Im}(D\Im _{k}\cdot \xi ^{\alpha })=q(k)-1$
which indeed sum up to $\dim _{\mathbb{R}}\,\mathcal{P}_{q(k)}=2q(k)-2$. $%
\blacksquare $

\bigskip

By projecting $\Im _{k}^{-1}(\{L_{\alpha }(k,t)\})$ into $\mathbb{CP}%
^{q(k)-2}$\ by means of $\pi :\mathcal{P}_{q(k)}\rightarrow \lbrack \mathcal{%
P}_{q(k)}]\simeq \mathbb{CP}^{q(k)-2}$\ we get the corresponding (connected
component of the) moduli space of polygons [10], 
\begin{equation}
\frak{MP}(\{L_{\alpha }(k,t)\})=\pi \left( \Im _{k}^{-1}(\{L_{\alpha
}(k,t)\})\right)
\end{equation}
parametrizing the equivalence classes, under similarities, of distinct
(possibly degenerate) polygons with the same edge-lengths $\{L_{\alpha
}(k,t)\}$ (and given fixed perimeter). Since in the neighborhood of regular
polygons $\Im _{k}^{-1}(\{L_{\alpha }(k,t)\})$ is a submanifold of $\mathcal{%
P}_{q(k)}$, the space $\Im _{k}^{-1}(\{L_{\alpha }(k,t)\})$ inherits from $%
\mathcal{P}_{q(k)}$ the structure of a principal $\mathbb{C}^{\ast }$-bundle
over the moduli space $\frak{MP}(\{L_{\alpha }(k,t)\})$, (with a singularity
structure, corresponding to degenerate polygons, which is described in
details in [10]).

As the edge-lengths $\{L_{\alpha }(k,t)\}$ vary in a neighborhood $U_{k}$ of 
$p_{eq}(k,a)$, we can characterize the corresponding space of polygons $\Im
_{k}^{-1}(\{L_{\alpha }(k,t)\})\cap U_{k}$ by means of the $1$-form field
defined, on the principal bundle $\mathcal{P}_{q(k)}$, by 
\begin{equation}
\psi (k)\doteq -\left[ a(\frac{\sqrt{3}}{3})q(k)\right] ^{-2}\sum_{\alpha
=1}^{q(k)-1}|Z^{\alpha }|\left( \sum_{\beta =1}^{\alpha }d|Z^{\beta
}|\right) ,  \label{unoforma}
\end{equation}
where the normalization to the fixed perimeter and the minus sign are chosen
for later convenience. As long as $\{L_{\alpha }(k,t)\}\neq \{0\}$, it
follows that $\ker \psi (k)=\ker D\Im _{k}$, so that $\Im
_{k}^{-1}(\{L_{\alpha }(k,t)\})$ are the integral distributions of $\psi (k)$%
. Such $\psi (k)$ is invariant under the involution (\ref{cconj}), whose
fixed point sets characterize degenerate polygons, and is well-defined also
if some of the $L_{\alpha }(k,t)\rightarrow 0$, (simply remove the
corresponding term from (\ref{unoforma})). Thus, it is easily checked that $%
\psi (k)$ extends to the whole space $\Im _{k}^{-1}(\{L_{\alpha }(k,t)\})$,
(degenerate and $L_{\alpha }(k,t)\rightarrow 0$ polygons included), and
provides a connection on $\Im _{k}^{-1}(\{L_{\alpha }(k,t)\})$ thought of as
a principal $\mathbb{C}^{\ast }$-fibration (with singularities) over the
moduli space of polygons $\frak{MP}(\{L_{\alpha }(k,t)\})$. The curvature $2$%
-form $\varpi (k)$ associated with such a connection can be locally written
as 
\begin{gather}
\varpi (k)=d\psi (k)=-\left[ a(\frac{\sqrt{3}}{3})q(k)\right]
^{-2}\sum_{\alpha =1}^{q(k)-1}d|Z^{\alpha }|\wedge \left( \sum_{\beta
=1}^{\alpha }d|Z^{\beta }|\right)  \label{tildeform} \\
=\left[ a(\frac{\sqrt{3}}{3})q(k)\right] ^{-2}\sum_{1\leq \alpha \leq \beta
\leq q(k)-1}d|Z^{\alpha }|\wedge d|Z^{\beta }|.  \notag
\end{gather}
It follows that the class $d\psi (k)$ represents the first Chern character
of the line bundle $\mathbb{\mathcal{L}}_{pol}(k)$ associated with the
fibration $\Im _{k}^{-1}(\{L_{\alpha }(k,t)\})\overset{\pi }{\rightarrow }%
\frak{MP}(\{L_{\alpha }(k,t)\})$,

\begin{equation}
c_{1}(\mathbb{\mathcal{L}}_{pol}(k))=\left\| d\psi (k)\right\| \in H^{2}(%
\frak{MP}(\{L_{\alpha }(k,t)\}),\mathbb{C}),
\end{equation}
where $H^{2}(\frak{MP}(\{L_{\alpha }(k,t)\}),\mathbb{C})$ denotes the second
deRham cohomology group of the polygons moduli space $\frak{MP}(\{L_{\alpha
}(k,t)\})$. Actually, $\left\| d\psi (k)\right\| $ lies in the image of the
inclusion $H^{2}(\frak{MP}(\{L_{\alpha }(k,t)\}),\mathbb{R})\subseteq H^{2}(%
\frak{MP}(\{L_{\alpha }(k,t)\}),\mathbb{C})$. Indeed, $d\psi (k)$ is the
pull-back, $d\psi (k)=\Im _{k}^{\ast }\omega (k)$,\ under the edge-length
map $\Im _{k}:\mathcal{P}_{q(k)}\longrightarrow \mathbb{R}^{q(k)}$, of the $%
\frak{S}_{q(k)}$- and scale-invariant $2$-form field defined on $\mathbb{R}%
^{q(k)-1}$ by 
\begin{equation}
\omega (k)=\left[ (\frac{\sqrt{3}}{3})q(k)\right] ^{-2}\sum_{1\leq \alpha
\leq \beta \leq q(k)-1}d\left( \frac{L_{a}(k)}{a}\right) \wedge d\left( 
\frac{L_{\beta }(k)}{a}\right)  \label{Kontform}
\end{equation}
providing the area elements (normalized to the fixed perimeter) associated
with the distinct pairs of\ bivectors $\frac{\partial }{\partial L_{\alpha }}%
\wedge \frac{\partial }{\partial L_{\beta }}$. Since 
\begin{equation}
dL_{\alpha }|_{a}=\frac{1}{3\sqrt{3}}\left[ dl_{\alpha }(k)-dl_{\alpha
+1}(k)+dl_{\alpha +2}(k)+dl_{\alpha ,\alpha +1}(k)+dl_{\alpha +1,\alpha
+2}(k)\right] ,  \label{polreg}
\end{equation}
we can also write (\ref{Kontform}), as 
\begin{equation}
\omega _{R}(k)=\left[ 3aq(k)\right] ^{-2}\sum_{1\leq \mu \leq \nu \leq
2q(k)}dl_{\mu }(k)\wedge dl_{\nu }(k),  \label{R2form}
\end{equation}
where $\{l_{\nu }(k)\}_{\nu =1}^{2q(k)}$ stand for the $2q(k)$ ($\nu $%
-relabelled) edge-lenghts of the simplicial loop of \ triangles $\{\Delta
_{\alpha }(k)\}$ which yields for the polygon $p(k)$ under dualization, (see
section 3). Note that for each given vertex $\sigma ^{0}(k)\in $ $%
|T_{l=a}|\rightarrow M$ the $2$-form (\ref{R2form}) allows for a complete
description of the infinitesimal deformations of $|T_{l=a}|\rightarrow M$
giving rise to a nearby Regge triangulation $|T_{L}|\rightarrow M$. It is
also worthwhile recalling that the $2$-form $\omega (k)$ naturally appears
in connection with certain universal polygonal bundles associated with the
combinatorial parametrization of the moduli space $\overline{\frak{M}}%
_{g},_{\lambda }$ introduced by Kontsevich in [1].

\subsection{\protect\bigskip Isoperimetric Regge measures}

In order to extend the above analysis to the deformations of the whole dual
polytope $|P_{T_{a}}|\rightarrow M$, associated with a dynamical
triangulation $|T_{l=a}|\rightarrow M$ with given curvature assignments, let
us consider the map, 
\begin{gather}
Def[\{p_{eq}(k,a),\delta _{k}\}_{k=1}^{N_{0}}]:\prod_{k=1}^{N_{0}(T)}[%
\mathcal{P}_{q(k)}]\longrightarrow \prod_{k=1}^{N_{0}(T)}\mathcal{P}_{q(k)}
\\
([p_{eq}(1,a)],...,[p_{eq}(k,a)],...)\longmapsto (p(1,\delta
_{1}),...,p(k,\delta _{k}),...)  \notag
\end{gather}
which associates with an ordered equivalence class of \ equilateral polygons 
$\{[p_{eq}(j,a)]\}_{j=1}^{N_{0}}$ a deformation realized by the polygonal $2$%
-cells of $\ $a polytope $|P_{T_{L}}|\rightarrow M$ obtained by
(isoperimetrically) varying the edge-lengths of $\ \delta _{k}\geq 0$ sides
of the corresponding equilateral $q(k)$-gon $\in |P_{T_{a}}|\rightarrow M$.
Clearly, not all possible deformations of the set of polygons $%
\{p_{eq}(k,a)\}_{k=1}^{N_{0}}$ give rise to a polytope $|P_{T_{L}}|%
\rightarrow M$. Since the number of edges of $|P_{T_{a}}|\rightarrow M$ is $%
N_{1}(T)$, and each varied edge is shared between two adjacent polygons, the
total number of varied edges is $\sum_{k=1}^{N_{0}(T)}2\delta _{k}$, and
must necessarily satisfy 
\begin{equation}
\sum_{k=1}^{N_{0}(T)}2\delta _{k}=N_{1}(T)-N_{0}(T),
\end{equation}
where the $N_{0}(T)$ comes from the isoperimetric constraints (\ref{lati}).
From Euler relation, we get 
\begin{equation}
\sum_{k=1}^{N_{0}(T)}\delta _{k}=N_{0}(T)+3g-3.  \label{deltakappa}
\end{equation}
Note that $\sum_{k=1}^{N_{0}(T)}2\delta _{k}$ also provides the maximal
dimension of the space of all possible polytopal deformations $%
Def[\{p_{eq}(k,a),\delta _{k}\}_{k=1}^{N_{0}}]$, i.e., 
\begin{equation}
\dim \left\{ Def[\{p_{eq}(k,a),\delta _{k}\}_{k=1}^{N_{0}}]\right\}
=2N_{0}(T)+6g-6,
\end{equation}
which, not surprisingly, coincides with the (real) dimension of the moduli
space $\frak{M}_{g},_{N_{0}}$ of Riemann surfaces of genus $g$ with $%
N_{0}(T) $ punctures. It is also important to stress that that the set of
deformation maps $\left\{ Def[\{p_{eq}(k,a),\delta
_{k}\}_{k=1}^{N_{0}}]\right\} $ is defined up to the action of the
automorphism group $Aut_{\partial }(P_{a})$ preserving the (labeling of the)
boundary components of $\ $the 1-skeleton$\ \Gamma $of the given dual
polytope $|P_{T_{a}}|\rightarrow M$. Thus,\ according to the remarks of
section 3, it easily follows that $\left\{ Def[\{p_{eq}(k,a),\delta
_{k}\}_{k=1}^{N_{0}}]\right\} $ has the structure of a differentiable
orbifold 
\begin{equation}
\left\{ Def[\{p_{eq}(k,a),\delta _{k}\}_{k=1}^{N_{0}}]\right\} \simeq \frac{%
\mathbb{R}_{+}^{N_{1}(T)-N_{0}(T)}}{Aut_{\partial }(P_{a})}.
\end{equation}
Since according to lemma 3 there is a one-to-one correspondance between the
polytopal deformations $Def[\{p_{eq}(k,a),\delta _{k}\}_{k=1}^{N_{0}}]$ and
the set of Regge triangulations which can be obtained by (isoperimetrical)
deformations from the given dynamical triangulation $|T_{l=a}|\rightarrow M$%
, the above remarks can be summarized in the following

\begin{proposition}
For any $|T_{l=a}|\rightarrow M$ $\in \mathcal{DT}[\{q(i)\}_{i=1}^{N_{0}}]$
the polytopal deformation space 
\begin{equation}
\frak{RG}[P_{T_{a}};\{q(i)\}_{i=1}^{N_{0}}]\doteq \bigcup_{\{\delta
_{k}\}}Def[\{p_{eq}(k,a),\delta _{k}\}_{k=1}^{N_{0}}]  \label{Reggedef}
\end{equation}
is naturally isomorphic to the space, $\frak{RG}[T;\{q(i)\}_{i=1}^{N_{0}}]$,
of all Regge triangulations $\{|T_{L}|\rightarrow M\}$, which can be
obtained by isoperimetrically deforming the equilateral polytopes $%
\{|P_{T_{a}}|\rightarrow M\}$ baricentrically dual to the given $%
|T_{l=a}|\rightarrow M$. \ Both spaces $\frak{RG}[P_{T_{a}};\{q(i)%
\}_{i=1}^{N_{0}}]$ and $\frak{RG}[T;\{q(i)\}_{i=1}^{N_{0}}]$ are
differentiable orbifolds modelled after $\frac{\mathbb{R}%
_{+}^{N_{1}(T)-N_{0}(T)}}{Aut_{\partial }(P)}$, of (maximal) dimension given
by 
\begin{equation}
\dim \frak{RG}[P_{T_{a}};\{q(i)\}_{i=1}^{N_{0}}]=\dim \frak{RG}%
[T;\{q(i)\}_{i=1}^{N_{0}}]=2N_{0}(T)+6g-6,
\end{equation}
and can be described in terms of the $2$- form 
\begin{equation}
\Omega (\frak{RG}[P_{T_{a}};\{q(i)\}_{i=1}^{N_{0}}])=\sum_{k=1}^{N_{0}(T)}%
\left[ (\frac{\sqrt{3}}{3}a)q(k)\right] ^{2}\varpi (k),  \label{Rvolum}
\end{equation}
where $\varpi (k)$ is given by (\ref{tildeform}).
\end{proposition}

Note that in describing deformations of dynamical triangulations, one can
use indifferently both $\frak{RG}[P_{T_{a}};\{q(i)\}_{i=1}^{N_{0}}]$ and $%
\frak{RG}[T;\{q(i)\}_{i=1}^{N_{0}}]$. However, form a mathematical point of
view, $\frak{RG}[P_{T_{a}};\{q(i)\}_{i=1}^{N_{0}}]$ is easier to handle
since it represents the natural choice for addressing the issue of the Regge
measure for the isoperimetric deformations of a given $|T_{l=a}|\rightarrow
M $.\ \ In this connection, the volume form naturally associated with (\ref
{Rvolum}) is 
\begin{equation}
\frac{\Omega (\frak{RG}[P_{T_{a}};\{q(i)\}])^{\dim \frak{RG}[P;\{q(i)\}]}}{%
\dim \frak{RG}[P_{T_{a}};\{q(i)\}]!},  \label{volum}
\end{equation}
(for notational simplicity we omit the superscripts in $\{q(i)%
\}_{i=1}^{N_{0}}$), it characterizes the set of all local (isoperimetric) $%
\{\delta _{k}\}_{k=1}^{N_{0}}$ Regge deformations around the given $T\in 
\mathcal{DT}[\{q(i)\}_{i=1}^{N_{0}}]$. Explicitly, we can write (\ref{volum}%
) as 
\begin{gather}
\frac{1}{(3g-3+N_{0}(T))!}\left( \sum_{k=1}^{N_{0}(T)}\left[ (\frac{\sqrt{3}%
}{3}a)q(k)\right] ^{2}\varpi (k)\right) ^{3g-3+N_{0}(T)}= \\
=\frac{1}{(3g-3+N_{0}(T))!}\left( \sum_{k=1}^{N_{0}(T)}\sum_{1\leq \alpha
\leq \beta \leq q(k)-1}d\left( L_{a}(k)\right) \wedge d\left( L_{\beta
}(k)\right) \right) ^{3g-3+N_{0}(T)}.  \notag
\end{gather}
If we relabel the edge lengths $\{\{L_{a}(k)\}_{\alpha
=1}^{q(k)}\}_{k=1}^{N_{0}(T)}\rightarrow \{L_{h}\}_{h=1}^{N_{1}(T)}$ we may
wonder about the connection of such a volume form with the Regge-like
measure on the set of all local deformations of $\{|P_{T_{a}}|\rightarrow
M\} $, \emph{i.e.}, 
\begin{equation}
dL_{1}\wedge dL_{2}\wedge ...\wedge dL_{N_{1}(T)},
\end{equation}
In order to address such a question, let us remark that if $\eta (k)$
denotes the perimeter of the generic polygon $p(k)$, then we can write 
\begin{equation}
\eta (k)\doteq \sum_{\alpha =1}^{q(k)}L_{a}(k)=\sum_{j}A_{(k)}^{j}L_{j}
\end{equation}
where $A_{(k)}^{j}$ is a $(0,1)$ indicator matrix with $A_{(k)}^{j}=1$ if
the edge associated with $L_{j}$ belongs to $p(k)$, and $0$ otherwise. Thus,
isoperimetric deformations, (\emph{i.e.}, $\eta (k)=(\frac{\sqrt{3}}{3}%
a)q(k) $), of the $N_{1}(T)$ edge lenghts $\{L_{j}\}$ are necessarily
subjected to the $N_{0}(T)$ linear constraints 
\begin{equation}
\left\{ \sum_{j}A_{(k)}^{j}L_{j}=(\frac{\sqrt{3}}{3}a)q(k)\right\}
_{k=1}^{N_{0}(T)},  \label{Lconstr}
\end{equation}
which, (since each edge is shared between two polygons), imply

\begin{align}
2\sum_{j=1}^{N_{1}(T)}L_{j}& =(\frac{\sqrt{3}}{3}a)%
\sum_{k=1}^{N_{0}(T)}q(k)=(\frac{\sqrt{3}}{3}a)b(n,n-2)|_{n=2}N_{0}(T)= \\
& =2\sqrt{3}a\left( N_{0}(T)+4g-4\right) ,  \notag
\end{align}
where $b(n,n-2)|_{n=2}$ is the average incidence of the dynamical
triangulation associated with $|P_{T_{a}}|\rightarrow M$, and where we have
used (\ref{avedue}). Thus, under the isoperimetric constraints (\ref{Lconstr}%
), the possible edge-lengths $\{L_{j}\}$ are necessarily restricted to a
subspace $\Delta (P_{T},\{q(k)\})$ of the $(N_{1}(T)-1)$-dimensional simplex 
\begin{equation}
\Delta _{N_{1}}(g,N_{0})\doteq \left\{ \{L_{j}\}\in \mathbb{R}%
_{+}^{N_{1}(T)}:\sum_{j=1}^{N_{1}(T)}L_{j}=\sqrt{3}a\left(
N_{0}(T)+4g-4\right) \,\right\} .
\end{equation}
The subspace $\Delta (P_{T},\{q(k)\})$ is the $(N_{1}-N_{0})$-dimensional
subsimplex of defined by 
\begin{gather}
\Delta (P_{T},\{q(k)\})\doteq \\
\left\{ \{L_{j}\}:\sum_{j=1}^{N_{1}(T)}L_{j}=\sqrt{3}a\left(
N_{0}(T)+4g-4\right) \,,\sum_{j}A_{(k)}^{j}L_{j}=(\frac{\sqrt{3}}{3}%
a)q(k)\right\} .  \notag
\end{gather}
Note that while $\Delta _{N_{1}}(g,N_{0})$ depends only from the fixed
parameter $N_{0}(T)$ and the genus $g$, the incidence structure of the
subsimplex $\Delta (P_{T},\{q(k)\})$ explicitly depends from the given
dynamical triangulation $|T_{a}|\rightarrow M$.

\bigskip

This is a convenient moment for recalling the following remarkable identity
proved by Kontsevich (lemma 3.1 and appendix C of [1]), 
\begin{gather}
\prod_{k=1}^{N_{0}(T)}d\eta (k)\wedge \left(
\sum_{k=1}^{N_{0}(T)}\sum_{1\leq \alpha \leq \beta \leq q(k)-1}d\left(
L_{a}(k)\right) \wedge d\left( L_{\beta }(k)\right) \right) ^{3g-3+N_{0}(T)}=
\label{kontid} \\
=2^{2N_{0}(T)+5g-5}(3g-3+N_{0}(T))!dL_{1}\wedge dL_{2}\wedge ...\wedge
dL_{N_{1}(T)},  \notag
\end{gather}
according to which we can write 
\begin{gather}
\frac{\Omega (\frak{RG}[P_{T_{a}};\{q(i)\}])^{\dim \frak{RG}[P;\{q(i)\}]}}{%
\dim \frak{RG}[P_{T_{a}};\{q(i)\}]!}\prod_{k=1}^{N_{0}(T)}d\eta (k,t)=
\label{omgauss} \\
=2^{2N_{0}(T)+5g-5}dL_{1}\wedge dL_{2}\wedge ...\wedge dL_{N_{1}(T)}.  \notag
\end{gather}
\ \ If we restrict the above volume form to the subsimplex $\Delta
(P_{T},\{q(k)\})$, then we immediately get the

\begin{lemma}
The volume form (\ref{volum}) is the Regge volume form restricted to the set
of Regge polytopes $\frak{RG}[P_{T_{a}};\{q(i)\}_{i=1}^{N_{0}}]$, i.e., 
\begin{gather}
\frac{\Omega (\frak{RG}[P_{T_{a}};\{q(i)\}])^{\dim \frak{RG}[P;\{q(i)\}]}}{%
\dim \frak{RG}[P_{T_{a}};\{q(i)\}]!}=  \label{Repol} \\
=2^{2N_{0}(T)+5g-5}\left. dL_{1}\wedge dL_{2}\wedge ...\wedge
dL_{N_{1}(T)}\right| _{\{\Delta (P,\{q(k)\})\}}  \notag
\end{gather}
\end{lemma}

Thus (\ref{volum}),\ (up to the fixed normalization factor), represents the
natural measure for the set of Regge polytopes that can be obtained by
isoperimetrically varying $\delta _{1}\geq 0$ edges of the $q(1)$-gone dual
to $\sigma ^{0}(1)\in |T_{l=a}|\rightarrow M$, then $\delta _{2}\geq 0$
edges of the $q(2)$-gone dual to $\sigma ^{0}(2)\in |T_{l=a}|\rightarrow M$,
and so on.

\subsection{Regge surfaces and 2D dynamical triangulations as moduli space
duals}

Since the space of polytopal deformations $Def[\{p_{eq}(k,a),\delta
_{k}\}_{k=1}^{N_{0}}]$ around any dynamical triangulation $T\in \mathcal{DT}%
[\{q(i)\}_{i=1}^{N_{0}}]$ is an orbifold locally modelled on $\mathbb{R}%
_{+}^{N_{1}(T)-N_{0}(T)}/Aut_{\partial }(P_{T})$,\ we can naturally define
an orbifold integration of any (sufficiently regular) functional $%
f(\{L_{\alpha }(k);T\})$ of the edge-lengths of a $|T_{l}|\rightarrow M\in 
\frak{RG}[T;\{q(i)\}_{i=1}^{N_{0}}]$ according to 
\begin{gather}
\int_{\mathcal{DT}[\{q(i)\}_{i=1}^{N_{0}}]}^{orb}f(\{L_{\alpha }(k);T\})
\label{orbint} \\
\doteq \sum_{T\in \mathcal{DT}[\{q(i)\}_{i=1}^{N_{0}}]}\frac{1}{%
|Aut_{\partial }(P_{T_{a}})|}\int_{_{\Delta (P,\{q(i)\})}}f(\{L_{\alpha
}(k);T\}),  \notag
\end{gather}
where the summation is over all distinct dynamical triangulations with given
curvature assignments weighted by the order $|Aut_{\partial }(P_{T})|$ of
the automorphisms group of the corresponding dual polytope. In particular,
we can integrate over $\mathcal{DT}[\{q(i)\}_{i=1}^{N_{0}}]$ the volume form
(\ref{volum}). In this way we can define the volume of all metrical
fluctuations around the set $\mathcal{DT}[\{q(i)\}_{i=1}^{N_{0}}]$ by means
of a duality pairing between Dynamical Triangulations and Regge polytopes, 
\begin{gather}
\left\langle \left\langle \mathcal{DT}[\{q(i)\}]\parallel \frak{RG}%
[P_{T_{a}};\{q(i)\}]\right\rangle \right\rangle \doteq  \label{topinv} \\
\doteq \int_{\mathcal{DT}[\{q(i)\}]}^{orb}\frac{\Omega (\frak{RG}%
[P_{T_{a}};\{q(i)\}])^{\dim \frak{RG}[P;\{q(i)\}]}}{\dim \frak{RG}%
[P_{T_{a}};\{q(i)\}]!}.  \notag
\end{gather}
Explicitly, by developing the integrand in (\ref{topinv}),\ we get 
\begin{gather}
\left\langle \left\langle \mathcal{DT}[\{q(i)\}]\parallel \frak{RG}%
[P_{T_{a}};\{q(i)\}]\right\rangle \right\rangle = \\
=\sum_{\{\delta _{i}:\tsum \delta
_{i}=2N_{0}(T)+6g-6\}}\prod_{i=1}^{N_{0}(T)}\frac{\left[ (\frac{\sqrt{3}}{3}%
a)q(i)\right] ^{2\delta _{i}}}{\delta _{i}!}  \notag \\
\times \left[ \sum_{T\in \mathcal{DT}[\{q(i)\}]}\frac{1}{|Aut_{\partial
}(P_{T_{a}})|}\int_{\Delta (P;\{q(i)\})}\prod_{k}^{N_{0}(T)}\varpi
(k)^{\delta _{k}}\right] ,  \notag
\end{gather}
\ which, according to the remarks of the previous section, can also be
written as 
\begin{gather}
\left\langle \left\langle \mathcal{DT}[\{q(i)\}]\left| \frak{RG}%
[P_{T_{a}};\{q(i)\}]\right. \right\rangle \right\rangle =  \label{Regint} \\
=\sum_{T\in \mathcal{DT}[\{q(i)\}]}\frac{2^{2N_{0}(T)+5g-5}}{|Aut_{\partial
}(P_{T_{a}})|}\int_{\Delta (P,\{q(i)\})}\left. dL_{1}\wedge dL_{2}\wedge
...\wedge dL_{N_{1}(T)}\right| _{\{\Delta (P,\{q(i)\})\}}.  \notag
\end{gather}
The integral 
\begin{equation}
\int_{\Delta (P,\{q(i)\})}\left. dL_{1}\wedge dL_{2}\wedge ...\wedge
dL_{N_{1}(T)}\right| _{\{\Delta (P,\{q(i)\})\}}=Vol\left[ \Delta
(P_{T},\{q(i)\})\right]
\end{equation}
is the volume of the $(N_{1}-N_{0})$-dimensional simplex $\Delta
(P_{T},\{q(i)\})$, and represents the Regge volume of the isoperimetrical
metrical fluctuations $Def[\{p_{eq}(k,a),\delta _{k}\}]$ around the
equilateral polytope $|P_{Ta}|\rightarrow M$ associated with the given
triangulation. Thus, we can write 
\begin{gather}
\left\langle \left\langle \mathcal{DT}[\{q(i)\}]\parallel \frak{RG}%
[P_{T_{a}};\{q(i)\}]\right\rangle \right\rangle = \\
=\sum_{T\in \mathcal{DT}[\{q(i)\}]}\frac{2^{2N_{0}(T)+5g-5}}{|Aut_{\partial
}(P_{T_{a}})|}Vol\left[ \Delta (P_{T},\{q(i)\})\right] =  \notag \\
=\sum_{\{\delta _{i}:\tsum \delta
_{i}=2N_{0}(T)+6g-6\}}\prod_{i=1}^{N_{0}(T)}\frac{\left[ (\frac{\sqrt{3}}{3}%
a)q(i)\right] ^{2\delta _{i}}}{\delta _{i}!}  \notag \\
\times \left[ \sum_{T\in \mathcal{DT}[\{q(i)\}]}\frac{1}{|Aut_{\partial
}(P_{T})|}\int_{\Delta (P,\{q(i)\})}\prod_{k}^{N_{0}(T)}\varpi (k)^{\delta
_{k}}\right] .  \notag
\end{gather}
The distinctive feature of \ the duality pairing (\ref{topinv})\ is that it
provides a connection between the measure of the isoperimetrical Regge
fluctuations and the topology of moduli space. In other words, the metrical
fluctuations in $\mathcal{DT}[\{q(i)\}]$ (represented by the distinct $%
|T_{l=a}|\rightarrow M$ which comprise such a space) and the Regge metrical
fluctuations represented by $Vol\left[ \Delta (P_{T},\{q(i)\})\right] $ are
not independent, but are related to the topology of the moduli space of
genus $g$ Riemann surfaces with $N_{0}$ punctures $\overline{\frak{M}}%
_{g},_{N_{0}}$. This remark is obviously a direct consequence of \ the fact
that our analysis of the interplay between dynamical triangulations and
Regge triangulations is strictly connected to Kontsevich's characterization
[1] of the intersection numbers $\left\langle \tau _{\delta _{1}}...\tau
_{\delta _{\lambda }}\right\rangle $ over the moduli space $\overline{\frak{M%
}}_{g},_{\lambda }$ (see (\ref{Kontnumb})). We get

\begin{proposition}
If $\left\langle \tau _{d_{1}}...\tau _{d_{N_{0}}}\right\rangle $ denote the
Witten-Kontsevich intersection numbers over the (compactified) moduli space
of genus $g$ Riemann surfaces with $N_{0}$ punctures$\overline{\frak{M}}%
_{g},_{N_{0}}$, then 
\begin{gather}
2^{2N_{0}(T)+5g-5}\sum_{T\in \mathcal{DT}[\{q(i)\}]}\frac{Vol\left[ \Delta
(P_{T},\{q(i)\})\right] }{|Aut_{\partial }(P_{T_{a}})|}=  \label{witKont} \\
=\sum_{\{\delta _{i}:\tsum \delta
_{i}=2N_{0}(T)+6g-6\}}\prod_{i=1}^{N_{0}(T)}\frac{\left[ (\frac{\sqrt{3}}{3}%
a)q(i)\right] ^{2\delta _{i}}}{\delta _{i}!}\left\langle \tau _{\delta
_{1}}...\tau _{\delta _{N_{0}}}\right\rangle .  \notag
\end{gather}
\end{proposition}

This immediately follows from the combinatorial description of $\ \overline{%
\frak{M}}_{g},_{N_{0}}$ in terms of ribbon graphs provided by Kontsevich
[1], (see also [5]). In particular, if \ $f:\frak{M}_{g},_{N_{0}}^{comb}%
\rightarrow \mathbb{R}_{+}^{N_{0}}$ denotes the projection which associates
to a (Jenkins-Strebel) ribbon graphs (\emph{i.e.,} the ribbon graph
associated with the Regge polytope $|P_{T_{L}}|\rightarrow M$) the sequence
of perimeters of the corresponding $q(k)$-gons, then according to [1] we
have 
\begin{equation}
\left\langle \tau _{\delta _{1}}...\tau _{\delta _{N_{0}}}\right\rangle
=\int_{f^{-1}(a(\frac{\sqrt{3}}{3})q(k))}^{orb}\prod_{k}^{N_{0}(T)}\widehat{%
\omega (k)}^{\delta _{k}},
\end{equation}
where \ $\int^{orb}$ denotes an orbifold integration similar to (\ref{orbint}%
), and where\ $\widehat{\omega (k)}$ is the pull-back of the $2$-form (\ref
{Kontform}) to$\overline{\frak{M}}_{g},_{N_{0}}\times \mathbb{R}_{+}^{N_{0}}$
under the map which assigns the edge-lengths to the boundary components of
Jenkins-Strebel differentials. It is easily verified that $\widehat{\omega
(k)}$ and $\varpi (k)$ are cohomologous and that by definition of orbifold
integration 
\begin{equation}
\int_{f^{-1}(a(\frac{\sqrt{3}}{3})q(k))}^{orb}\prod_{k}^{N_{0}(T)}\widehat{%
\omega (k)}^{\delta _{k}}=\sum_{T\in \mathcal{DT}[\{q(i)\}]}\frac{1}{%
|Aut_{\partial }(P_{T_{a}})|}\int_{\Delta
(P,\{q(i)\})}\prod_{k}^{N_{0}(T)}\varpi (k)^{\delta _{k}},
\end{equation}
from which the stated result follows. $\blacksquare $

\bigskip

Thus, we get 
\begin{gather}
2^{2N_{0}(T)+5g-5}\sum_{T\in \mathcal{DT}[\{q(i)\}]}\frac{Vol\left[ \Delta
(P_{T},\{q(i)\})\right] }{|Aut_{\partial }(P_{T_{a}})|}=  \label{Inter} \\
=(\frac{1}{3}a^{2})^{N_{0}+3g-3}F_{g}\left( q(1),q(2),...\right) ,  \notag
\end{gather}
where 
\begin{equation}
F_{g}\left( q(1),q(2),...\right) \doteq \sum_{\{\delta
_{i}\}}\prod_{i=1}^{N_{0}(T)}\frac{\left[ q(i)\right] ^{2\delta _{i}}}{%
\delta _{i}!}\left\langle \tau _{\delta _{1}}...\tau _{\delta
_{N_{0}}}\right\rangle ,  \label{genfunct}
\end{equation}
is the exponential generating function of the intersection numbers on $%
\overline{\frak{M}}_{g},_{N_{0}}$, [6], [1]. From (\ref{Inter}) we get a
non-trivial connection between the Euclidean volumes of the simplices $%
\Delta (P_{T},\{q(i)\})$ and intersection theory on $\overline{\frak{M}}%
_{g},_{N_{0}}$ Since $\overline{\frak{M}}_{g},_{N_{0}}$ is locally modelled
on the combinatorial orbifold $\mathbb{R}_{+}^{N_{1}(T)-N_{0}(T)}/Aut_{%
\partial }(P_{T})$, it is natural to interpret the Regge volumes 
\begin{equation}
\frac{Vol\left[ \Delta (P_{T},\{q(i)\})\right] }{|Aut_{\partial }(P_{T_{a}})|%
}
\end{equation}
as the the volumes of the open strata in $\overline{\frak{M}}_{g},_{N_{0}}$,
each stratum being labelled by the distinct \ \ dynamical triangulations $%
T\in \mathcal{DT}[\{q(i)\}]$. Such an interpretation gives to the (seemingly
trivial) isoperimetrical Regge measure (\ref{Repol}) a deep geometrical
meaning which, a priori, is totally unexpected. From a physical point of
view the relation (\ref{Inter}) also shows that the average volume \ \ $%
\left\langle Vol\left[ \Delta (P_{T},\{q(i)\})\right] \right\rangle $, (the
average being over the discrete set $\mathcal{DT}[\{q(i)\}]$), is such that 
\begin{equation}
\left\langle Vol\left[ \Delta (P_{T},\{q(i)\})\right] \right\rangle Card%
\left[ \mathcal{DT}[\{q(i)\}_{i=1}^{N_{0}}]\right] =\frac{(\frac{1}{3}%
a^{2})^{N_{0}+3g-3}}{2^{2N_{0}(T)+5g-5}}F_{g}\left( q(1),q(2),...\right) ,
\label{duty}
\end{equation}
where $Card\left[ \mathcal{DT}[\{q(i)\}_{i=1}^{N_{0}}]\right] $ denotes the
number of distinct dynamical triangulations with given curvature assignments 
$\in \mathcal{DT}[\{q(i)\}_{i=1}^{N_{0}}]$. This latter cardinality
comprises an important part of the total triangulations counting, \emph{i.e.}%
, 
\begin{equation}
Card\left[ \mathcal{DT}[N_{0}]\right] =\sum_{\{q(i)\}_{i=1}^{N_{0}}}Card%
\left[ \mathcal{DT}[\{q(i)\}_{i=1}^{N_{0}}]\right] ,
\end{equation}
where the summation extends over all possible curvature assignments \ for
triangulations (with $N_{0}$ vertices) over a surface of given topology, (%
\emph{e.g.}, see [11] and references cited therein), and where $\mathcal{DT}%
[N_{0}]$ denotes the set of distinct dynamical triangulations with $N_{0}(T)$
marked vertices admitted by a surface of genus $g$. It is well known that,
for large $N_{0}(T)$, we get 
\begin{equation}
Card\left[ \mathcal{DT}[N_{0}]\right] \sim N_{0}(T)^{\gamma
_{g}+N_{0}-3}e^{\mu _{0}N_{0}}\left( 1+O(\frac{1}{N_{0}})\right) ,
\label{asympt}
\end{equation}
where 
\begin{equation}
\gamma _{g}\doteq \frac{5g-1}{2},
\end{equation}
is the genus-$g$ pure gravity critical exponent, and $\mu _{0}$ is a
(non-universal) parameter independent of $g$ and $N_{0}$. Thus, (\ref{duty})
expresses a topological duality between a relevant part of the natural
counting measure for dynamical triangulation theory and the
(isoperimetrical) Regge measure one naturally associates with Regge
(polytopal) surfaces. It tells us that intersection theory on $\overline{%
\frak{M}}_{g},_{N_{0}}$ \ (hence topological 2D gravity) comes from a
balance between the entropy $Card\left[ \mathcal{DT}[\{q(i)\}_{i=1}^{N_{0}}]%
\right] $ and the typical Regge deformation volume $\left\langle Vol\left[
\Delta (P_{T},\{q(i)\})\right] \right\rangle $. Roughly speaking, this
reflect the fact that the smaller the typical volume of the strata the
higher the number of representative points \ \ $|T_{l=a}|\rightarrow M$
needed in order to combinatorially approximate \ $\overline{\frak{M}}%
_{g},_{N_{0}}$. If we define 
\begin{equation}
A[T;\{q(j)\}]\doteq \ln Vol\left[ \Delta (P_{T},\{q(i)\})\right] ,
\end{equation}
then (\ref{Inter}) implies that 
\begin{equation}
\sum_{T\in \mathcal{DT}[N_{0}]}\frac{1}{|Aut_{\partial }(P_{T_{a}})|}%
e^{A[T;\{q(j)\}]}=\frac{(\frac{1}{3}a^{2})^{N_{0}+3g-3}}{2^{2N_{0}(T)+5g-5}}%
\sum_{\{q(i)\}_{i=1}^{N_{0}}}F_{g}\left( q(1),q(2),...\right) .
\end{equation}
Thus, the free energy $A[T;\{q(j)\}]$ associated with the Regge measure
plays against the entropic factor $Card\left[ \mathcal{DT}[N_{0}]\right] $
so as to give rise to what is basically a topological quantity. From a
physical point of view, such a\emph{\ }mechanism, \emph{a la Peierls},\emph{%
\ }indicates that 2D dynamical triangulations and Regge calculus are related
by a form of topological $S$-duality. Clearly the physical relevance of this
latter remark is biased by the fact that we are explicitly considering here
an isoperimetric family of Regge measures and its relation with the standard
measure [12], [13] it is not clear. (Notice however that\ the isoperimetric
constraint can be removed by integrating over the perimeters $\{\eta (k)\}$
still mantaining, according to Kontsevich's lemma, a Regge measure
structure). One may \ also stress the fact that, from a deformation theory
point of view, we get (\ref{volum})\ as a unique choice. Perhaps, this
unicity might help in shedding light on the controversial issue of the most
appropriate choice of a measure in quantum Regge calculus and related models
(see,\emph{\ e.g.}, [14]). Another challenging open problem is to exploit
such a duality to compute the scaling behavior of $Vol\left[ \Delta
(P_{T},\{q(i)\})\right] $ and the associated critical exponents. This would
provide an interesting way of estimating volumes of open strata in $%
\overline{\frak{M}}_{g},_{N_{0}}$,

\bigskip

\textbf{Acknowledgements}

This work was supported in part by the Ministero dell'Universita' e della
Ricerca Scientifica under the PRIN project \emph{The geometry of integrable
systems.}

\bigskip

\bigskip

\textbf{References}

\bigskip

[1] M. Kontsevich, \emph{Intersection theory on moduli space of curves},
Commun. Math. Phys. \ \ \ 147, \ (1992) 1.

[2] M. Troyanov, \emph{Prescribing curvature on compact surfaces with
conical singularities}, Trans. Amer. Math. Soc. 324, (1991) 793; see also M.
Troyanov, \emph{Les surfaces euclidiennes a' singularites coniques},
L'Enseignment Mathematique, 32 (1986) 79.

[3] W. P. Thurston, \emph{Shapes of polyhedra and triangulations of the
sphere}, Geom. and Topology Monog. Vol.1 (1998) 511.

[4]E. Picard, \emph{De l'integration de l'equation }$\Delta u=e^{u}$\emph{\
sur une surface de Riemann fermee'}, Crelle's Journ. 130 (1905) 243.

[5]M. Mulase, M. Penkava, \emph{Ribbon graphs, quadratic differentials on
Riemann surfaces, and algebraic curves defined over }$\overline{\mathbb{Q}}$%
, math-ph/9811024 v2 (1998).

[6]E. Witten, \emph{Two dimensional gravity and intersection theory on
moduli space}, Surveys in Diff. Geom. 1. (1991) 243.

[7] K. Strebel, \emph{Quadratic differentials}, Springer Verlag, (1984).

[8] V. A. Voevodskii, G.B. Shabat, \emph{Equilateral triangulations of
Riemann surfaces, and curves over algebraic number fields}, Soviet Math.
Dokl. 39 (1989) 38.

[9] M. Bauer, C. Itzykson, \emph{Triangulations}, in: \emph{The Grothendieck
Theory of Dessins d'Enfants}, ed. L. Schneps, Lond. Math. Soc. Lect. Notes
Series, Vol. 200, Cambridge Univ. Press (1994) 179.

[10] M. Kapovich, J. Millson, \emph{On the moduli space of polygons in the
Euclidean plane},

J. Diff. Geom. 42 (1995) 133.

[11] J. Ambj\o rn, B. Durhuus, T. Jonsson, \emph{Quantum Geometry},
Cambridge Monograph on \ \ Mathematical Physics, Cambridge Univ. Press
(1997).

[12] H. Hamber, R. Williams, \emph{On the measure in simplicial gravity},
Phys. Rev. D 59 (1999) 064014-1.

[13] H. Hamber, \emph{On the gravitational scaling dimensions},
hep-th/9912246.

[14] P. Menotti, P. Peirano,\emph{\ Functional integration on
two-dimensional Regge geometries}, Nucl. Phys. B 473 (1996) 426, see also
Phys. Lett. B 353, \ (1995) 444.

\end{document}